\documentclass[letterpaper,11pt]{article}
\pdfoutput=1 

\usepackage{jheppub} 

\usepackage{tikz}
\usepackage{dsfont}
\usepackage{amsmath}
\def\ba{\begin{align}}\def\ea{\end{align}}

\definecolor{darkgreen}{rgb}{0,0.4,0}
\def\beq{\begin{eqnarray}}\def\eeq{\end{eqnarray}}
\def\be{\begin{equation}}\def\ee{\end{equation}}
\def\ben{\begin{equation}}
\def\een{\end{equation}}
\def\bea{\begin{eqnarray}}
\def\eea{\end{eqnarray}}

\def\vx{{\vec{x}}}

\def\cO{{\cal{O}}}

\def\vev#1{\langle{#1}\rangle}

\def\tr{\rm tr}

\def\t6 {T_\mt{D6}}

\newcommand{\mt}[1]{\textrm{\tiny #1}}

\newcommand{\bQ}{{\bar{Q}}}
\newcommand{\bx}{{\bar{x}}}

\def\cale         {{\cal E}}

\def\ee           {{\rm e}}

\def\tr           {\mathop{\rm Tr}}

\def\sqr#1#2{{\vcenter{\vbox{\hrule height.#2pt
 \hbox{\vrule width.#2pt height#1pt \kern#1pt
 \vrule width.#2pt}\hrule height.#2pt}}}}

\def\e{\epsilon}

\def\ee{\cale}

\def\aa1{\phi}
\def\cc1{\psi}

\def\nn{\nabla_\nu}

\def\vev#1{\langle #1 \rangle}

\def\vx{\vec{x}}

\def\comment#1{{\bf [[#1]]}}
\def\tw{\tilde{w}}

\def\bra#1{{\langle}#1|}
\def\ket#1{|#1\rangle}

\def\vev#1{\langle{#1}\rangle}

\def\nd{{ \vphantom{\dagger}}}

\def\eg{{\it e.g.}}
\def\ie{{\it i.e.}}

\def\etal{{\it et. al.}}
\title{Exactly Solvable Floquet Dynamics for Conformal Field Theories in Dimensions Greater than Two}

\author{Diptarka Das $^1$,}
\author{Sumit R. Das$^2$,}
\author{Arnab Kundu$^{3,4}$,}
\author{Krishnendu Sengupta $^5$,}

\affiliation{$^1$ Department of Physics, Indian Institute of Technology, Kanpur, UP 208016, INDIA.}
\affiliation{$^2$Department of Physics and Astronomy, University of Kentucky, Lexington, KY 40506, U.S.A.}
\affiliation{$^3$Saha Institute of Nuclear Physics, 1/AF Bidhannagar, Kolkata 700064, INDIA.}
\affiliation{$^4$Homi Bhaba National Institute, Training School Complex, Anushaktinagar, Mumbai 400094, INDIA.}
\affiliation{$^5$School of Physical Sciences, Indian Association for the Cultivation of Science, Jadavpur, Kolkata 700032, INDIA.}

\emailAdd{didas@iitk.ac.in}
\emailAdd{das@pa.uky.edu}
\emailAdd{arnab.kundu@saha.ac.in}
\emailAdd{ tpks@iacs.res.in}

\abstract{We find classes of driven conformal field theories (CFT) in $d+1$ dimensions with $d > 1$, whose quench and floquet dynamics can be computed exactly. The setup is suitable for studying periodic drives, consisting of square pulse protocols for which Hamiltonian evolution takes place with different deformations of the original CFT Hamiltonian in successive time intervals. These deformations are realized by specific combinations of conformal generators with a deformation parameter $\beta$; the $\beta < 1$ ($\beta > 1$) Hamiltonians can be unitarily related to the standard (L\"uscher-Mack) CFT Hamiltonians.  The resulting time evolution can be then calculated by performing appropriate conformal transformations. For $d \leq 3$ we show that the transformations can be easily obtained in a quaternion formalism. Evolution with such a single Hamiltonian yields qualitatively different time dependences of observables depending on the value of $\beta$, with exponential decays characteristic of heating for $\beta > 1$, oscillations for $\beta < 1$ and power law decays for $\beta =1$. This manifests itself in the behavior of the fidelity, unequal-time correlator, and the energy density at the end of a single cycle of a square pulse protocol with different hamiltonians in successive time intervals. When the Hamiltonians in a cycle involve generators of a single $SU(1,1)$ subalgebra we calculate the Floquet Hamiltonian. We show that one can get dynamical phase transitions for any $\beta$ by varying the time period of a cycle, where the system can go from a non-heating phase which is oscillatory as a function of the time period to a heating phase with an exponentially damped behavior.  Our methods can be generalized to other discrete and continuous protocols. We also point out that our results are expected to hold for a broader class of QFTs that possesses an SL$(2, C)$ symmetry with fields that transform as quasi-primaries under this. As an example, we briefly comment on celestial CFTs in this context.}

\begin{document}

\begin{flushright}
\end{flushright}

\maketitle
\flushbottom
\vspace{10pt}

\section{Introduction}

The study of time evolution of a driven quantum system with a time dependent Hamiltonian is a valuable tool for gaining insight into its non-equlibrium properties. Useful drive protocols include quantum quench, ramp, and periodic or quasi-periodic drive protocols \cite{rev1,rev2,rev3,rev4}. Recently, properties of quantum systems driven via periodic protocols have been most
intensely studied; such systems undergo
evolution governed by Hamiltonian periodic in time with a time period $T$. Interestingly, at strobosocopic times $t=nT$, where $n$ is an integer, the evolution operators of such systems can be written as $U(nT,0)= \exp[-i H_F n T]$; the corresponding dynamics is then completely controlled by the Floquet Hamiltonian $H_F$.

However, most of these studies require numerical work: analytically tractable models are limited to free field theories or two dimensional conformal field theories. For the latter class of theories, the underlying infinite dimensional symmetry algebra provides a powerful tool to calculate physical quantities of interest. One example involves a sudden quench from a massive theory to a two dimensional CFT whose IR properties can be well approximated by a Cardy-Calabrese state \cite{ccstate}, and the time evolution can be calculated by using conformal mapping. A more recent example involves quench and Floquet dynamics in a two dimensional CFT \cite{Wen:2018vux, Goto:2021sqx, Goto:2023wai,  Kudler-Flam:2023ahk, Goto:2023yxb, Nozaki:2023fkx,Wen:2018agb, Lapierre:2019rwj, Wen:2020wee, Fan:2020orx, Das:2021gts}: once again the time evolution can be obtained by conformal maps. No such analytical results are known in higher dimensions, and one has to typically resort to numerical calculations which are often limited by finite size effects \footnote{A different class of problems involving fast smooth quenches in conformal field theories in arbitrary dimension can be addressed using conformal properties \cite{dasmyers,dymarsky}. }.

In this paper we demonstrate that a class of time dependent problems can be exactly studied in conformal field theories in {\em arbitrary number of dimensions} and lead to interesting non-trivial dynamics. Our work is inspired by the work of \cite{Wen:2018vux} who studied a $1+1$  dimensional CFT on a strip which starts from the usual CFT Hamiltonian and is quenched by a sine squared deformed (SSD) one. The SSD-CFT${}_2$ has been studied previously in \cite{Ishibashi:2016bey, Tada:2014kza, Okunishi:2016zat, Tada:2017wul, Caputa:2020mgb}. Both these Hamiltonians belong to the global part of the Virasoro algebra. The time evolution then becomes a M\"obius transformation. The global part of the Virasoro algebra is (in Euclidean signature) $SO(3,1)$ (generated by $L_0, L_{\pm 1}$ and their hermitian conjugates). Floquet generalizations to  periodically driven CFTs by a sequence of the usual CFT Hamiltonian and SSD Hamiltonian for successive time durations $T_i$ were studied in \cite{Wen:2018agb, Wen:2020wee, Das:2021gts}. Generalizations of the SSD to other $SL_2$ subgroups of Virasoro were studied in \cite{Fan:2020orx}. Holographic interpretations were investigated in \cite{Goto:2021sqx, Goto:2023wai,  Kudler-Flam:2023ahk, Nozaki:2023fkx} {\color{black} and from a different perspective in \cite{Das:2022pez}}.

In higher dimensions ($d+1$), the conformal algebra is $SO(d+2,1)$. This suggests that a natural generalization to higher dimensional systems can be obtained by considering different generators of the conformal algebra as Hamiltonians and considering sequences of non-commuting Hamiltonians in successive time intervals. Similar to the two dimensional case, the dynamics is expected to be equivalent to conformal transformations.

In this work we consider conformal field theories on $S^d \times {\rm time}$. The class of Hamiltonians are combinations of generators which belong to $SL(2,R) \sim SU(1,1)$ subrgoups of the conformal group. More specifically we consider
\ben
H (\beta, \Pi) = 2iD + i\beta (K_\mu + P_\mu) \Pi^\mu \ ,
\label{1-1}
\een
where the operators $D,K_\mu,P_\mu$ ($\mu,\nu = 0, \cdots d$) are generators of conformal transformations (see below) and $\Pi^\mu$ is a projector along $\mu$-direction (e.g. $\Pi^0 = (1,0,0,..)$). { In the field theory Hilbert space the hermiticity properties of these generators are
\ben
D^\dagger = - D~~~~~~~~K_\mu^\dagger = - P_\mu
\label{hermiticity}
\een}{black}
Note that there are $(d+1)$ such different SL$(2,R)$ subalgebras corresponding to $(d+1)$-inequivalent choices of $\Pi^\mu$. This provides a wider class of drive protocols where one can use different members of the class of Hamiltonians $H(\beta, \Pi)$, for different time intervals. { In the latter case, the net transformation is not a $SU(1,1)$ transformation.}

In terms of the energy momentum tensor, the Hamiltonian on $S^d \times {\rm{(time)}}$ may be written as
{\color{black}
\ben
H(\beta, \Pi) = 2\int d\Omega_{d}~[1 + \frac{1}{2}\beta Y_\mu \Pi^\mu]  T_{ww}\\
\label{1-2a}
\een
where $Y^\mu$ denote cartesian coordinates on a $R^{d+1}$, where the sphere $S^d$ is the surface $Y^\mu Y_\mu =1$, $d\Omega_{d}$ denotes the volume element on $S^d$ and $w$ denotes the Euclidean time.}

As will be shown below, when $\beta < 1$ a Hamiltonian of the form (\ref{1-1}) can be transformed to the standard  CFT Hamiltonian $(H_{\rm ST}=2iD)$ (on the plane), while for $\beta > 1$ it can be transformed to the L\"uscher-Mack (LM) type Hamiltonian $(H_{\rm LM}=i(P_\mu + K_\mu))$. We note that {\color{black}for a CFT on $S^d$},  $H_{\rm LM}$ shares the same vacuum as the $H_{\rm ST}$. However as we shall show, the two Hamiltonians are related by a non-unitary transformation. Furthermore unitary time evolution with $H_{\rm LM}$ results in exponential time dependence.


As is standard in CFT's, it is convenient to first work in Euclidean signature and perform a Weyl transformation to a plane $R^{d+1}$
\begin{eqnarray}
ds^2 &=& dw^2 + d\Omega_{d}^2 = \frac{1}{r^2} \left[\delta_{\mu\nu}dx^\mu dx^\nu \right] \label{planecylinder1}
\end{eqnarray}
where $x^\mu$ are cartesian coordinates on $R^{d+1}$ and
\begin{eqnarray} 
r^2 &=& \delta_{\mu\nu} x^\mu x^\nu = e^w. \label{planecylinder2}
\end{eqnarray} 
The operator $(2iD)$ becomes the dilatation operator on the plane, while $P_\mu$ generates translations on the plane.
Acting on functions on $R^{d+1}$ the conformal generators are represented by the following differential operators
\begin{eqnarray}
D &=& -i x_{\mu} \partial_{\mu}, \quad  P_{\mu} = -i
\partial_{\mu}, \quad K_{\mu} = -i ( 2 x_{\mu} ( x_{\nu} \partial_{\nu}) -
r^2 \partial_{\mu}),  \nonumber \\
L_{\mu\nu} &=& -i (x_\mu \partial_\nu - x_\nu \partial_\mu) \ .
\label{gen1}
\end{eqnarray}
For a fixed $\mu$ the generators $D,P_\mu,K_\mu$ form an SU$(1,1)$ subalgebra of the {\color{black}Euclidean} conformal algebra SO$(d+2,1)$. There
are $d+1$ such subgroups corresponding to different choice of $\mu$. The ${\rm SU}(1,1)$ nature of these subgroups can be understood by noting
that the commutation of these generators satisfy
\begin{eqnarray}
[D, K_{\mu}] &=& -i K_{\mu}, \quad  [D, P_{\mu}] = i P_{\mu}, \quad
[K_{\mu}, P_{\mu}] = 2i D \ . \label{gencom}
\end{eqnarray}

Our strategy for calculating the response is analogous to the earlier works in $1+1$ dimensions. We will perform a Weyl transformation to $R^{d+1}$, where the expressions for the generators and the resulting conformal transformation become simple, calculate the time evolution by evaluating the corresponding conformal transformation, transform back to $S^d \times {\rm time}$ and finally continue to real time.
While this is straightforward in principle, this becomes quickly cumbersome in practice for $d > 1$. 

The key technical tool which facilitates our calculations is the fact that when $d\le 3$, a point on $R^{d+1}$ can be represented by a quaternion and the action of finite conformal transformations take a simple form, generalizing M\"obius transformations on the complex plane to M\"obius transformations on the field of quaternions, $SL(2,H)$. In general, the parameters of these transformations are themselves quaternions. However, when the transformation belongs to an $SL(2,R)$ subgroup of the conformal group one can judiciously choose the quaternion representation such that these parameters become real numbers.  When continued to real time, these transformations are represented by $SU(1,1)$ transformations on quaternions and correspond to real transformations of the coordinates
While this simplifies our calculations considerably, this necessitates switching quaternionic representations whenever we switch from one $SL(2,R)$ subgroup to another. It should be emphasized that when a cycle involves different $SL(2,R)\sim SU(1,1)$ subgroups the net transformation is not a 
$SL(2,R)\sim SU(1,1)$ transformation. This is then essentially different from the $d=1$ case investigated in the literature.
In this work, we provide explicit results for $3+1$ dimensions, though it should be emphasized that the framework is completely general for any $d\le 3$.

In the following we first consider properties of various quantities after a single drive cycle as a function of the time period $T$ of the drive and $\beta$.  Each cycle involves Hamiltonians of the form (\ref{1-1}) with various $\Pi^\mu, \mu = 0,1,\cdots3$ for successive time intervals of {\em equal} value. The value of $\beta$ is either zero, or some fixed value, i.e. we do not consider different non-zero values of $\beta$ for different time intervals. These restrictions are for simplicity.
All these diagonistics show that for $\beta < 1$ the response is oscillatory as a function of the time {\color{black} period}, while for $\beta > 1$ one exponential decays for large time {\color{black} period}. $\beta = 1$ is a critical value, where the quantities have power law behavior. Since the driving Hamiltonians break rotation invariance on the $S^d$ the system develops inhomogenities. The nature of the inhomogeneties is richer in these higher dimensional cases, since we have the ability to break different sets of symmetries by choosing different sequences of $H_1 \cdots H_{d+1}$.
It can be easily seen that if we use different nonzero values of $\beta$ the behavior is oscillatory when all the $\beta$ values are less than 1, power law when all the $\beta$ are equal to one. If any of the $\beta$ exceed 1, the behavior will be exponential.

{\color{black}
When all the hamiltonians in a cycle belong to the same $SU(1,1)$ subgroup of the conformal group, we calculate the Floquet hamiltonian explicitly, so that the behavior after an arbitrary number of cycles, $n$, can be then read off easily. We show that the three conjugacy classes of $SU(1,1)$ correspond to exponential, oscillatory and power law behaviors as a function of the number of cycles, $n$. By changing the time intervals inside a single cycle we show that one can make transitions between these different phases both for $\beta < 1$ and $\beta > 1$. }

When the cycles involve different Mobius subgroups, one needs to look at conjugacy classes of $SL(2,H)$. We indicate how one can proceed in this general case, but defer a detailed investigation to a future publication.

The plan of the rest of this work is as follows. In Sec.\ \ref{deformed} we study the properties of Hamiltonians of the form (\ref{1-1}). This is followed by Sec.\ \ref{dynamics} which deals with our strategy for calculating the response to the dynamics for general $d$ and the formulation in terms of quaternions for $d=3$. Next, in Sec.\ \ref{results}, we provide results for three quantities under the dynamics. The first corresponds to the fidelity of an evolving primary state at the end of a drive cycle; this is discussed in subsection (\ref{fidsec}). The second constitutes
behavior of unequal-time correlation functions of the CFT under such evolution; this is discussed subsection(\ref{corrsec}). The
evolution of the stress tensor, starting from a primary CFT state, is discussed in subsection (\ref{stressec}). { As expected, the conformal anomaly plays a role in this calculation.} {\color{black} In section (\ref{multiple}) we find the fixed points and fixed surfaces for quaternionic M\"obius transformations which belong to a single $SU(1,1)$ and determine the trajectories of points under successive cycles of a periodic drive}. Finally, we discuss our main results and conclude in Sec.\ \ref{conc}. Some details of the calculation are provided in the appendix.

\section{The deformed CFTs}
\label{deformed}

In this section we will discuss some properties of deformed CFT's with a Hamiltonian of the form (\ref{1-1}).

\subsection{$\beta < 1$ and M\"obius quantization}

When $\beta < 1$, for the choice e.g. $\Pi^\mu= (1, 0,0,0)$, the Hamiltonian $H$ (defined in (\ref{1-1})) can be thought of a different quantization of the theory with the undeformed hamitonian ($\beta = 0$) upto a scaling of the new time. For $d=1$ this has been called ``M\"obius quantization''. Consider the Hamiltonian
\ben
{\tilde H} = \frac{1}{\sqrt{1-\beta^2}} H =( \cosh \theta ) 2iD + i( \sinh \theta) (K_0+P_0)~~~~~~~\beta = \tanh \theta \ .
\label{3-1}
\een
This Hamiltonian is related to the dilatation operator by a similarity transformation
\begin{eqnarray}
U^{-1} {\tilde H} U^{} &=& (2iD), \quad  U= \exp \left[-\frac{i}{2}\theta (K_0-P_0) \right] \ .
\label{trans1_1}
\end{eqnarray}
This can be verified by explicit calculation. However since the left hand side of (\ref{trans1_1}) depends only on the commutators of the generators of the $SU(1,1)$ group, the equation must be independent of the specific representation.

To verify (\ref{trans1_1}) it is convenient to use a representation of the $SU(1,1)$ using Pauli matrices,
\begin{eqnarray}
D &=&  i \sigma_z/2, \quad K_{0} = \sigma_-, \quad P_{0} =
\sigma_+ \quad \sigma_\pm = \frac{1}{2}(\sigma_x \pm  i \sigma_y) . \label{rep1}
\end{eqnarray}
The transformed special conformal and translation generators can be now deduced using the identities
\begin{eqnarray}
U^{-1} \sigma_z U & \equiv & \tau_z=  \sigma_z \cosh \theta +  i
\sigma_x \sinh \theta \nonumber\\
U^{-1} \sigma_x U & \equiv & \tau_x =  \sigma_x \cosh \theta - i \sigma_z
\sinh \theta, \quad  \tau_y = \sigma_y \ . \label{utrans11}
\end{eqnarray}
{Using the representation (\ref{rep1}) this leads to
\begin{eqnarray}
U^{-1} D U &=& D \cosh \theta + \frac{1}{2} (K_0 +P_0) \sinh \theta
\nonumber\\
U^{-1} K_0 U &=& \frac{1}{2}(1+\cosh \theta) K_0 - \frac{1}{2}
(1-\cosh \theta) P_0  - D \sinh \theta \ , \nonumber \\
U^{-1} P_0 U &=& \frac{1}{2}(1+\cosh \theta) P_0 - \frac{1}{2}
(1-\cosh \theta) K_0  - D \sinh \theta \ .
\label{genrel1}
\end{eqnarray}}
To calculate the transformation of the other generators of the conformal algebra, it is useful to consider the combinations
\ben
A_\mu \equiv \frac{1}{2}(K_\mu - P_\mu)~~~~~~~~~B_\mu \equiv \frac{1}{2}(K_\mu + P_\mu) \ .
\label{3-2}
\een
The conformal algebra then leads to a closed $SL(2,R)$ subalgebra for the generators $(A_0, A_j, L_{0j})$ for each value of $j=1,2,3$
\ben
[A_0,L_{0j}]=i A_j~~~~~~[A_0,A_j]=iL_{0j}~~~~~~[L_{0j},A_j] =iA_0 \ .
\label{3-4}
\een
Now consider the transformed $A_j$ or $L_{0j}$ (these angular momentum generators are defined in (\ref{gen1})),
{
\ben
U^{-1} A_j U= e^{i\theta A_0}A_j e^{-i\theta A_0}~~~~~~
U^{-1} L_{0j} U = e^{i\theta A_0}L_{0j} e^{-i\theta A_0}
\label{3-3}
\een
}
Once again, the result is determined entirely in terms of commutators. We can therefore use any representation of the algebra (\ref{3-4}) to perform the calculation, e.g. the representation in terms of Pauli matrices
\ben
L_{0j} \rightarrow \frac{1}{2} \sigma_x~~~~A_j \rightarrow \frac{1}{2}i\sigma_z~~~~~A_0 \rightarrow -\frac{1}{2}i\sigma_y
\een
to obtain
{
\bea
U^{-1} A_j U & = & A_j \cosh \theta - L_{0j} \sinh \theta \nonumber \\
U^{-1} L_{0j} U& = & L_{0j} \cosh \theta - A_j \sinh \theta \ .
\label{3-5}
\eea
}
Furthermore the conformal algebra implies
\ben
[ A_0 , K_j + P_j ] = [A_0, L_{ij}] = 0 \ .
\label{3-6}
\een
{This leads to final form of the deformed generators:
\begin{eqnarray}
U^{-1} D U  &=& D \cosh \theta -\frac{1}{2} (K_0 +P_0) \sinh \theta
\nonumber\\
U^{-1} K_0 U &=& \frac{1}{2}(1+\cosh \theta) K_0  - \frac{1}{2}
(1-\cosh \theta) P_0 - D \sinh \theta \nonumber\\
U^{-1} P_0 U&=& \frac{1}{2}(1+\cosh \theta) P_0  - \frac{1}{2} (1-\cosh
\theta) K_0 - D \sinh \theta \nonumber\\
U^{-1} K_{j} U &=& \frac{1}{2}(1+\cosh \theta) K_j + \frac{1}{2} (1-\cosh
\theta) P_j - L_{0j} \sinh \theta \nonumber\\
U^{-1} P_{j} U &=& \frac{1}{2}(1+\cosh \theta) P_j + \frac{1}{2} (1-\cosh
\theta) K_j + L_{0j} \sinh \theta  \nonumber\\
U^{-1} L_{0j} U &=& \cosh \theta \, L_{0j} + \frac{1}{2}\sinh \theta (P_j-K_j), \quad
U^{-1} L_{ij} U= L_{ij} \ .  \label{allgen}
\end{eqnarray}}

These relations can be alternatively derived by first looking at the coordinate transformations resulting from the action of $U$ with infinitesimal $\theta$, and exponentiating them and requiring that the correct commutation relations are satisfied by the deformed generators. This is detailed in the appendix.

The results of this subsection imply that for $\beta < 1$ the deformed Hamiltonian is proportional to a standard CFT Hamiltonian (dilatation operator) which is quantized with a different notion of time. Consequently aspects of the physics of the deformed Hamiltonian are expected to be qualitatively similar to the undeformed Hamiltonian.

\subsection{$\beta > 1$ and L\"uscher-Mack Hamiltonians}

The unitary transformation which relates the deformed theory to the undeformed theory for $\beta < 1$ does not work when $\beta > 1$. We will now demonstrate that for $\beta > 1$ the Hamiltonian can be instead deformed to the generator $K_0 + P_0$. Consider the Hamiltonian
{
\ben
{\hat{H}} = \frac{1}{\sqrt{\beta^2-1}}H = 2iD \sinh \phi + i(K_0+P_0)\cosh \phi~~~~\beta = \coth \phi \ .
\label{4-1}
\een
Using manipulations entirely similar to the previous subsection it is easy to show that
\ben
U^{-1}{\hat{H}}U = i(K_0+P_0),\quad  U= \exp \left[-\frac{i}{2}\phi (K_0-P_0) \right] \ .
\label{trans2}
\een}
The transformations of the other generators can be worked out following a procedure entirely analogous to the previous subsection.
This shows that the physics of the deformed theory for $\beta > 1$ is similar to a theory with a L\"uscher-Mack Hamiltonian $H_{LM} = \frac{i}{2} (K_0+P_0)$ rather than the dilatation operator. { In fact one has the following relationship: 
\begin{align}
 \frac{i}{2} ( K + P)  &= e^{ \tfrac{\pi}{4} ( K - P ) } D  e^{ -\tfrac{\pi}{4} ( K - P ) } .\label{eq:kpd} 
\end{align}
If we let $\ket{\tilde{\Delta}_n}$ be the eigenstate of $D$, \ie $\,\,D \ket{\tilde\Delta_n} = \tilde\Delta_n \ket{\tilde\Delta_n}$ (with $\tilde\Delta_n = i \mathbb{R}$), then one can define left and right eigenvectors of LM Hamiltonian: 
\begin{align}
\ket{n_r } &= e^{ \tfrac{\pi}{4} \left( K - P \right) } \ket{\tilde\Delta_n} ,\,\,\,
\bra{n_\ell } = \bra{\tilde\Delta_n} e^{ -\tfrac{\pi}{4} \left( K - P \right) }.
\end{align}
These states satisfy: $\vev{ n_\ell | n'_r } = \delta_{\tilde\Delta^\nd_n,\tilde\Delta'_n}$ and, $\sum\limits_n  \ket{n_r } \bra{n_\ell}  = \mathbb{I}$. 
This construction is similar to the one recently used in \cite{Chen:2023hra}. If we consider the thermal partition function of the LM theory at temperature $T$, which is :  $Z_{LM}(T) = \tr e^{ - \frac{i}{2 T} \left( K + P \right) }$, we may use the above eigenstates to show that this is given by $\tr \,\,e^{- \frac{i}{T}  \left(iD\right) }$. Hence finite temperature for LM theory gets Wick-rotated into imaginary temperature partition function of the standard CFT: 
\begin{align}
Z_{LM}( T) &= Z_{CFT}( -i T ) . 
\end{align}
This is suggestive of the fact that if we look at real time evolution in LM, we may have thermal like behaviours marked by exponential dependence in real time. This we can confirm using the $2 \times 2$ representations Eq.\eqref{rep1}. The unitary evolution operator $U(t)$ takes the form: 
\begin{align}
U(t) &= \exp \left( -i t H_{LM} \right) = e^{\frac{t}{2} \sigma_x } = \begin{pmatrix}
\cosh \frac{t}{2} & \sinh \frac{t}{2} \\
\sinh \frac{t}{2} & \cosh \frac{t}{2}
\end{pmatrix}. 
\end{align}
}

As we will see, this difference manifests in dynamical processes like quantum quench by the appearance of a heating phase for $\beta > 1$ and an oscillating phase for $\beta < 1$.

\section{Dynamics as a Conformal Transformation}
\label{dynamics}

As mentioned above, it is convenient to think of the time evolution by Hamiltonians like (\ref{1-1}) by first performing a Weyl transformation to $R^{d+1}$ using Eqs.\ \ref{planecylinder1} and \ref{planecylinder2}.
Euclidean time evolution with the Hamiltonian $H(\beta, \Pi)$ is equivalent to a conformal transformation. This transformation can be obtained by using the Baker-Campbell-Hausdorff formula
\cite{Martinez-Tibaduiza:2020brq}
\begin{eqnarray}
U_{\mu}\Pi^\mu \equiv U(\Pi) &=&   e^{-(2i w D + i w \beta (P_\mu + K_\mu) \Pi^\mu)} =
e^{\Lambda_+ K_{\mu} \Pi^\mu} e^{\ln \Lambda_0 i D/2} e^{\Lambda_- P_{\mu} \Pi^\mu}
\nonumber\\
\Lambda_0 &=& \left(\cosh w \nu_0 -(w \nu_0)^{-1} \sinh w\nu_0
\right)^{-2}, \nonumber\\
\Lambda_{+} &=& -\Lambda_-= i \beta \nu_0^{-1} \sinh(w\nu_0) \Lambda_0^{1/2}
\label{bch1}
\end{eqnarray}
where we have defined
\ben
\nu_0= \sqrt{1-\beta^2}
\label{nuzero}
\een
To write down the explicit transformations, we separate out the components. For example, using the above relations, one obtains:
\begin{eqnarray}
e^{\alpha_1 iD} x_{\nu} e^{-\alpha_1 iD}&=& e^{\alpha_1} x_{\nu}, \quad  e^{i\alpha_2
P_{\mu}} x_{\nu}e^{-i\alpha_2
P_{\mu}} = x_{\nu} + \alpha_{2}\hat{x}_\mu, \nonumber\\
e^{i\alpha_3 K_{\mu}} x_{\nu} e^{-i\alpha_3 K_{\mu}} &=&
\frac{x_{\nu}}{1 - 2 \alpha_3 x_{\mu} + r^2 \alpha_3^2} \ , \label{bch2}
\end{eqnarray}
where $K_{\mu}$ and $x_{\mu}$ denote the $\mu$-th component of the corresponding vectors.\footnote{To present these formulae in a completely covariant form, one needs to implement an appropriate projection operator, which we do not use here. Our notation therefore breaks covariance. Nonetheless, we hope it is clear from the context whether we are referring to a vector or a particular component of it.} Identifying $\alpha_1= (1/2) \ln \Lambda_0$,
$\alpha_2= -i \Lambda_-$, and $\alpha_3= -i \Lambda_+$, we find,
after a few lines of algebra and for $\nu \ne \mu$
\begin{align}
&x'_{\nu} = \frac{x_{\nu}}{{\cal{D}}_{\mu}}, \label{trans1} \\
&\text{where the denominator, } {\cal{D}}_{\mu} = \Big[\left(\cosh w\nu_0 -(w \nu_0)^{-1} \sinh w\nu_0
\right)^2 + 2 x_{\mu} \beta \nu_0^{-1} \sinh w \nu_0 \nonumber\\
&\times \left(\cosh w \nu_0
- (w \nu_0)^{-1} \sinh w \nu_0 \right)
+ \beta^2 r^2 (w \nu_0)^{-2} \sinh^2
\tau \nu_0 \Big]. \nonumber
\end{align}
A similar, but more complicated expression can be obtained for
$\nu=\mu$ in a similar manner. One obtains
\begin{eqnarray}
x'_{\mu} &=& \Big[ x_{\mu}(2(\cosh^2 w \nu_0 -(w \nu_0)^{-2} \sinh^2 w \nu_0)-1) - \beta \nu_0^{-1} \sinh w \nu_0 (\cosh w \nu_0 (1-r^2) \nonumber\\
&& -
(w \nu_0)^{-1} \sinh w \nu_0 (1+r^2)) \Big]/{\cal{D}}_{\mu} \ . \label{trans2}
\end{eqnarray}
These finite conformal transformations can be then used to express time evolved quantities in terms of the quantities at initial time for any $d$.

\subsection{$d=3$ and quaternions}
\label{qtn} 

In {\color{black} $d \le 3$} dimensions in general an efficient way to compute the
transformed coordinates under the $SU(1,1)$ subgroup of conformal transformations is to use the
quaternion formulation. In what follows, we provide details this formulation for $d=3+1$.
In Euclidean signature the coordinates are denoted by $X^\mu$ with $x_0= \tau$, $x_1=x$ and
so on. We shall also define the $2 \times 2$ matrices $\tau_0= I$
and $\tau_j = -i \sigma_j$ where $\sigma_j$ are the standard Pauli
matrices and $I$ denotes the identity matrix. The first step of
using the quaternion formulation is then to write the coordinates
$x_{\mu}$ using a $2 \times 2$ matrix by associating each component
of the coordinate to one of the $\tau_{\mu}$, where $\vec \tau= (I,
-i \sigma_j)$, for $j=1,2,3$:
\begin{eqnarray}
Q_{\nu} &=& I x_{\nu} - i \sum_{j=1}^3 \sigma_{j} y_{j} \ . \label{q1}
\end{eqnarray}
The $y_j$ which appears here are the components of $x^\mu$ with $\mu \neq \nu$.
The choice of $\nu$ and $\mu$
is arbitrary at this stage. As an example, we may choose $\nu=1$. The three other coordinates $x^j$ which appear are $y_1=x_0=\tau, y_2=x_2=y, y_3=x_3=z$, leading to
\begin{eqnarray}
Q_{1} &=& \left( \begin{array}{cc} x- iz & -i (\tau-iy) \\ -i(\tau+ i
y) & x+iz \end{array} \right)  \label{q2}
\end{eqnarray}
Clearly $Q_{\nu}$ is not unique even after we fix $\nu$ and
associate $x_{\nu}$ with the identity matrix since there is freedom
of associating the rest of the coordinates with other Pauli matrices
in different ways. {All such choices lead to
identical results for final coordinates under the class of conformal
transformations that we discuss. Moreover it is also possible  
to carry out the transformation when $Q$ is chosen in a different manner, {\it i.e.}, 
without necessarily associating $x_{\nu}$ with the identity matrix; this is discussed in details in \S App.\ \ref{appb}.}

A general conformal transformation on $R^{d+1}$ is then represented by a quaternionic Mobius transformation
\ben
Q'_{\mu} = I x'_{\mu} - i \sum_{j \neq \mu}\sigma_{j} x'_{j} = ({\tilde a}_1 Q_{\mu} - i
{\tilde a}_2 I). (i {\tilde a}_3 Q_{\mu} + {\tilde a}_4 I)^{-1} 
\label{quatermobius}
\een
where ${\tilde a}_1,{\tilde a}_2,{\tilde a}_3,{\tilde a}_4$ are quaternions obeying
\ben
Det \left[{\tilde a}_1 {\tilde a}_3^{-1}{\tilde a}_4 {\tilde a}_3 - {\tilde a}_2 {\tilde a}_3 \right] = 1
\label{unitdeterminant}
\een
{\color{black}
Two successive transformations lead to another Mobius transformation where the matrices $a_i$ get multiplied by matrix multiplication \cite{qua3}. Consider transforming $Q'_\mu$ by a further mobius transformation,
\ben
Q^{\prime\prime}_{\mu}  = I x''_{\mu} - i \sum_{j \neq \mu}\sigma_{j} x''_{j}
= ({\tilde a}_1' Q'_{\mu} - i
{\tilde a}_2' I). (i {\tilde a}_3' Q'_{\mu} + {\tilde a}_4' I)^{-1} 
\label{quatermobius2}
\een
Then 
\ben
Q^{\prime\prime}_{\mu}  = ({\tilde a}_1'' Q_{\mu} - i
{\tilde a}_2''I). (i {\tilde a}_3'' Q_{\mu} + {\tilde a}_4'' I)^{-1} 
\label{qua2}
\een
where
\bea
{\tilde a}_1'' & = & {\tilde a}_1'{\tilde a}_1+ {\tilde a}_2'{\tilde a}_3~~~~~~~{\tilde a}_2''
={\tilde a}_1'{\tilde a}_2+ {\tilde a}_2'{\tilde a}_4 \nonumber \\
{\tilde a}_3'' & = & {\tilde a}_3' {\tilde a}_1+{\tilde a}_4' {\tilde a}_3~~~~~~~{\tilde a}_4''
={\tilde a}_3'{\tilde a}_2+{\tilde a}_4' {\tilde a}_4
\label{composition}
\eea
}

However, for the $SL(2,R)\sim SU(1,1)$ subgroup generated by $D,K_\mu,P_\mu$ these parameters become real numbers provided one chooses the quaternion $Q_\mu$ as in (\ref{q1}), i.e. with $x^\mu$ being the coefficient of the identity matrix. The action of the SU(1,1) transformations is then obtained by representing the generators by Pauli matrices as in (\ref{rep1}). The operator $U(\Pi^\mu)$ in (\ref{bch1}) is then represented by the $2 \times 2$ matrix,
\begin{eqnarray}
U(\Pi^\mu) &=&   e^{-w(-\sigma_z + i \beta \sigma_x)} =\left(
\begin{array}{cc} \tilde{a}_1 & \tilde{a}_2 \\ \tilde{a}_3 & \tilde{a}_4 \end{array} \right)
\nonumber\\
\tilde{a}_1 &=& \left(\cosh w\nu_0 + \nu_0^{-1} \sinh w\nu_0 \right), \quad
\tilde{a}_4= \left(\cosh w\nu_0 - \nu_0^{-1} \sinh w\nu_0 \right) \nonumber\\
\tilde{a}_2 &=& \tilde{a}_3 = -i \beta \nu_0^{-1} \sinh w\nu_0 \label{utrans1}
\end{eqnarray}
with $\tilde{a}_1 \tilde{a}_4-\tilde{a}_2 \tilde{a}_3=1$. To find the transformed coordinates, we
note that the quaternion matrix $Q_{\mu}$ transforms, upon action of
$U(\Pi^\mu)$ to $Q'_{\mu}$ given by \cite{qua1, qua2, qua3, Pal:2020dqf}
\begin{eqnarray}
Q'_{\mu} &=& I x'_{\mu} - i \sum_{j \neq \mu}\sigma_{j} x'_{j} = (\tilde{a}_1 Q_{\mu} - i
\tilde{a}_2
I). (i\tilde{a}_3 Q_{\mu} + \tilde{a}_4 I)^{-1} \nonumber\\
&=& \left(
\begin{array}{cc} \frac{ x_{\mu}(2\tilde{a}_1 \tilde{a}_4-1) -i (\tilde{a}_2 \tilde{a}_4 - \tilde{a}_1 \tilde{a}_3 r^2) -
ix_{\nu_1}}{\tilde{a}_4^2 + 2 i \tilde{a}_3 \tilde{a}_4 x_{\mu} - \tilde{a}_3^2 r^2} & \frac{-i
(x_{\nu_2}- i x_{\nu_3})}{\tilde{a}_4^2 + 2 i \tilde{a}_3 \tilde{a}_4 x_{\mu} - \tilde{a}_3^2 r^2} \\
\frac{-i (x_{\nu_2}+ i x_{\nu_3})}{\tilde{a}_4^2 + 2 i \tilde{a}_3 \tilde{a}_4 x_{\mu} -
\tilde{a}_3^2 r^2} & \frac{x_{\mu}(2\tilde{a}_1 \tilde{a}_4-1) -i (\tilde{a}_2 \tilde{a}_4 - \tilde{a}_1 \tilde{a}_3 r^2) +
ix_{\nu_1}}{\tilde{a}_4^2 + 2 i \tilde{a}_3 \tilde{a}_4 x_{\mu} - \tilde{a}_3^2 r^2}
\end{array} \right) \label{qtrans1}
\end{eqnarray}
where $ \mu \neq \nu_1$, $\nu_2$, and $\nu_3$. The
transformed coordinates are then obtained from the relation
$x'_{\alpha} = {\rm Tr}[\tau_{\alpha}^{-1} Q'_{\mu}]/2$ and yields
\begin{eqnarray}
x'_{\mu} &=& \frac{ x_{\mu}(2\tilde{a}_1 \tilde{a}_4-1) -i (\tilde{a}_2 \tilde{a}_4 - \tilde{a}_1 \tilde{a}_3
r^2)}{\tilde{a}_4^2 + 2 i \tilde{a}_3 \tilde{a}_4 x_{\mu} - \tilde{a}_3^2 r^2}, \quad x'_{\nu} =
\frac{ x_{\nu}}{\tilde{a}_4^2 + 2 i \tilde{a}_3 \tilde{a}_4 x_{\mu} - \tilde{a}_3^2 r^2}
\label{qtrans2}
\end{eqnarray}
Substituting Eq.\ \ref{utrans1} in Eq.\ \ref{qtrans2}, one recovers
Eqs.\ \ref{trans1} and \ref{trans2}. This shows that the quaternionic
approach provides a vastly simpler way to obtain the transformed
coordinates.

In the rest of the paper we will deal with $d=3$. The transformation from $R \times S^3$ to $R^4$ is given by
\begin{eqnarray}
\tau &=& e^w \cos \theta \ ,  x= e^w \sin \theta \cos \phi \ ,
y= e^w \sin \theta \sin \phi \cos \psi \ , z= e^w \sin \theta \sin\phi \sin \psi  \ , \label{tran1}
\end{eqnarray}
where $w_{\mu}=(w,\theta,\phi,\psi)$ are the coordinates on the
cylinder $R \times S^3$, and $w$ denotes time on the cylinder.

{ The above transformations are in euclidean signature. Upon continuation to real time $ w \rightarrow \tw = -iw$ while the angles remain the same. Let us denote the analytically continued transformation matrix by
\ben
U = \left(
\begin{array}{cc} a & b \\ c & d \end{array} \right)
\een
For the transformation (\ref{utrans1}) we have $a = d^\star$ while $b=c^\star = {\rm real}$. Using (\ref{qtrans2}) it may be checked that the transformed time, $\tw^\prime$ and the transformed angles $\theta^\prime, \phi^\prime, \psi^\prime$ remain real.

We will be interested in time evolution with square pulse protocol with different $\beta$ and different $\Pi_\mu$ in different intervals. When these hamiltonians have the {\em same} $\Pi_\mu$ the time evolution in the different intervals correspond to transformations which lie in the {\em same $SU(1,1)$ subgroup} of the conformal group. In this case the net transformation can be simply obtained by usual matrix multiplication of the different $U$'s. The net parameters will obey $a=d^\star$ and $b = c^\star$, which are the $SU(1,1)$ conditions, but $b,c$ will not be in general real. However, the $SU(1,1)$ conditions are sufficient to ensure that the transformed coordinates on $(time) \times S^3$ are real.

The expression (\ref{utrans1}) shows that the behavior of the system under real time evolution depends on $\beta$. Continuing to real time ${\tilde w}$ and recalling that $\nu_0$ is given by (\ref{nuzero}) and we see that that the resulting conformal transformation involves trigonometric functions of time for $\beta < 1$, hyperbolic functions for $\beta > 1$ and power laws when $\beta = 1$. Thus for late times, observables will decay exponentially in time for $\beta > 1$ and oscillate when $\beta < 1$. }

While we have illustrated this formalism for $d=3$ the same formalism can be used for $d=1,2$ by setting some of the coordinates in Eq.\eqref{q2} to zero.

\section{Floquet Dynamics}
\label{results}

In this section, we shall study Floquet dynamics of $3+1$
dimensional CFTs, using a square pulse protocol, described below. 
{\color{black} The protocols we consider have different hamiltonians in successive time intervals and we choose these time intervals to be equal.}
We will compute various physical properties at the end of each cycle. The strategy can be generalized to arbitrary number of cycles. We defer the analysis for multiple cycles to future work. In what follows, we shall choose the amplitude
of the pulses within a drive period (denoted by $\beta$ below) to be same for simplicity. 

 In most of this section, we shall focus on dynamics of driven system at the end of a drive cycle; more specifically, we shall study the behavior of the operator expectation and fidelity starting from a primary state and unequal time correlators starting from vacuum, at $t=T$ as a function of $T$ and $\beta$. {\color{black}  Here $T$ is the time extent (or time period) of a single cycle.} Our analysis can be extended to study the micromotion (dynamics for $t \le T$) of these quantities; however, in the present work we focus on their properties at $t=T$. 

 The corresponding macromotion, which yields information about stroboscopic dynamics of the driven system at $t=nT$ ($n \in Z$ and $n>1$),
is governed by the Floquet Hamiltonian $H_F$; we shall derive $H_F$ for a class of drive protocols involving a single  $SU(1,1)$ subgroup in Sec.\ \ref{flha}.
{ In section (\ref{multiple}) we discuss the structure of fixed points and surfaces for this special case.} However, the corresponding analysis of macromotion for a generic drive protocol which involves multiple $SU(1,1)$ subgroups is more complicated and is left as a subject of future study.

The strategy will be to perform the computations on $R^4$ (coordinates $x^\mu$), Weyl transforming to $R \times S^3$ (coordinates $w^\mu$), and finally analytically continuing to real time $t = iw$.
Under the Weyl transformation, a primary operator with conformal dimension $\Delta$ transforms as
\begin{eqnarray}
O_{\Delta}(w_\mu) &=& e^{w \Delta} O_\Delta(x_\mu). \label{tran1}
\end{eqnarray}
We shall use these relations to transform between the physical
coordinates $w_{\mu}$ on the cylinder and those on the plane
($x_{\mu}$).

As discussed above, unitary time evolution governed by a Hamiltonian of the form (\ref{1-1}) is equivalent to a conformal transformation $x_\mu \rightarrow x_\mu^\prime $ given by (\ref{tran1}).
The transformation of the primary operators of
the CFT having conformal dimension $\Delta$ due to such dynamics is
given by \cite{Simmons-Duffin:2016gjk} :
\begin{eqnarray}
O(x_{\mu}) &\rightarrow& U^{\dagger} O(x_{\mu}) U = O(x'_{\mu})
J_2^{\Delta/4}, \quad J_2 = \left | \frac{\partial
x'_{\mu}}{\partial x_{\nu}} \right| = |{\rm Det}[ i \tilde{a}_3 Q + \tilde{a}_4
I]|^{-4}\label{optran1}
\end{eqnarray}
where the last expression holds if the coordinates ${\bf x}'$ and
${\bf x}$ are related by the transformation given by Eqs.\
\ref{utrans1} and \ref{qtrans1}.

\begin{figure}
\centering
\rotatebox{0}{\includegraphics*[width= 0.55 \linewidth]{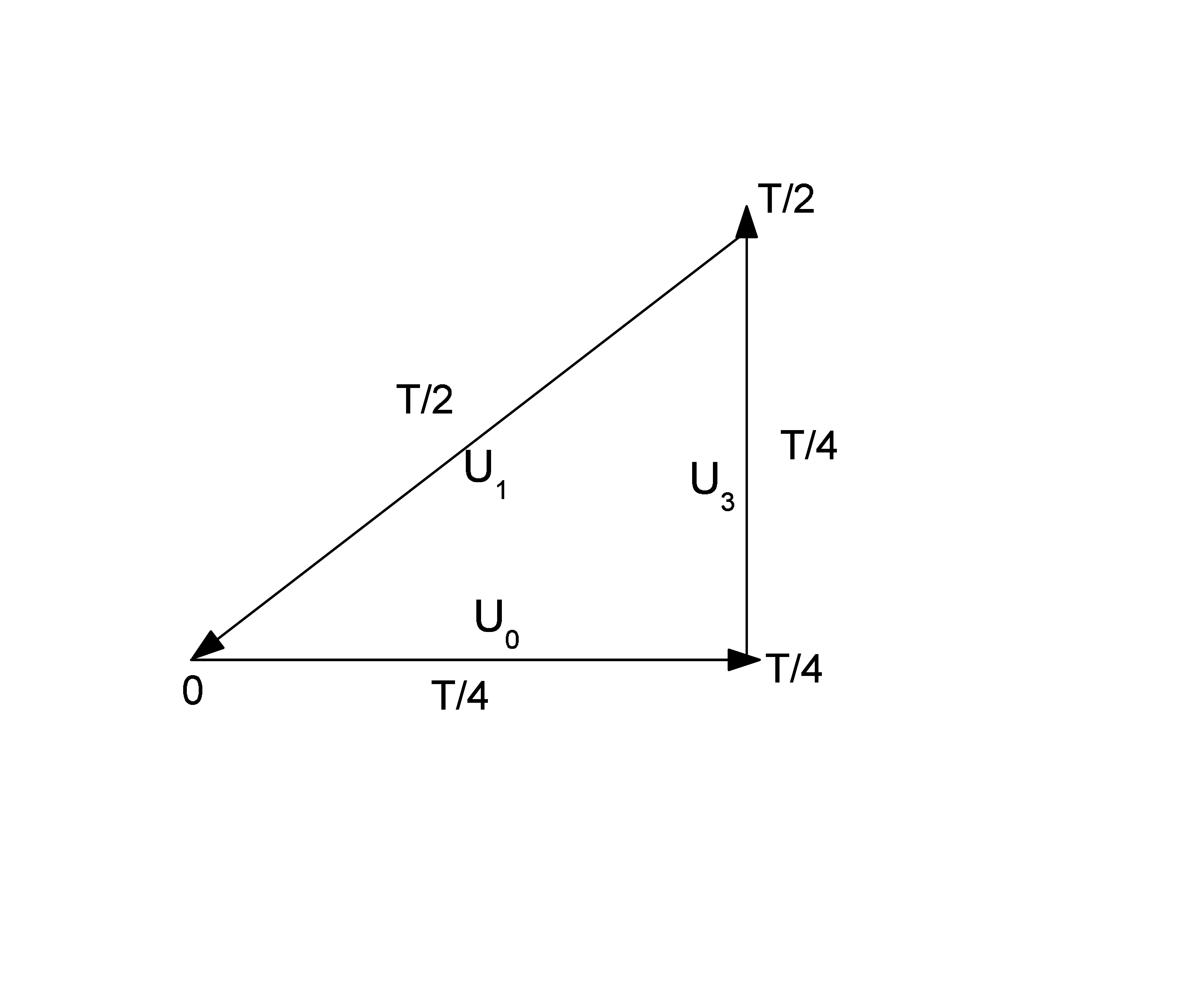}}
\caption{Schematic representation of the protocol for computing
fidelity $F(t)$. See text for details. \label{fig1}}
\end{figure}

\subsection{Fidelity}
\label{fidsec}

In this section, we shall compute the fidelity $F(T)$ of a primary
state $|\Delta\rangle$ of the CFT at the end of a drive cycle. This is
defined, in Euclidean time, as
\begin{eqnarray}
F(T_0) &=& \frac{ \langle \Delta | U(T_0,0)|\Delta\rangle}{\langle \Delta
|\Delta \rangle} = \frac{ \lim_{x_{2 \mu} \to \infty, x_{1 \mu} \to 0}
\langle 0| \phi({\bf x}_2) U(T_0,0) \phi({\bf x}_1)
|0\rangle}{\lim_{x_{2 \mu} \to \infty, x_{1 \mu} \to 0} \langle
0|\phi({\bf x}_2) \phi({\bf x}_1) |0\rangle} \label{fid1}
\end{eqnarray}
where $\phi({\bf x})\equiv \phi(\tau,x,y,z)$ denotes a primary CFT
field of dimension $h$, $T=-iT_0$ is the drive period in real time
and $|0\rangle$ denotes the CFT vacuum. Throughout this section, we
shall work in Euclidean time and analytically continue to real time
whenever necessary.

The protocol we choose for computing $F(T)$ is schematically shown
in Fig.\ \ref{fig1}. The evolution operator $U(\Pi^{\mu})$ in each cycle
is chosen to be
\begin{eqnarray}
U(\Pi^{\mu}; \tau,0) &=& e^{-H(\Pi^{\mu}) \tau}, \quad  H(\Pi^{\mu} )= i( 2D
+ \beta (K_{\mu} + P_{\mu}) \Pi^\mu \ . \label{evol1}
\end{eqnarray}
The total evolution
operator at the end of one cycle is given by
\begin{eqnarray}
&& U = U^{\dagger}(\Pi^1 ;0,T_0/2) U(\Pi^3; T_0/2,T_0/4) U(\Pi^0; T_0/4,0) \ , \\
&& \Pi^1=(0,1,0,0)\ , \quad \Pi^3=(0,0,0,1) \ , \quad \Pi^0=(1,0,0,0) \ .
\label{evol2}
\end{eqnarray}
Note that we use the different deformed CFT Hamiltonians with same
deformation parameter $\beta$ to generate the evolution operator
$U$. The fidelity is computed at the end of the cycle.

To obtain $F(T)$, we first note that the two point correlation
function of a primary operator with dimension $\Delta$ can be written in the quaternion formalism as \cite{qua3}
\begin{eqnarray}
C_0 &=& \langle 0| \phi({\bf x}_2) \phi({\bf x}_1) |0\rangle =
\frac{1}{({\rm Det}[Q({\bf x}_2)-Q({\bf x}_1)])^\Delta} \label{cexp1}
\end{eqnarray}
In the above equation we have not specified a subscript for the quaternions since this particular result is independent of which $Q_{(\mu)}$ we use.
For the unequal-time correlation function under a transformation by
$U_{\mu}(T,0)$, this leads to
\begin{eqnarray}
C_1 &=& \langle 0|U_{\mu}^{\dagger}(T,0)\Pi^\mu \phi({\bf x}_2)
U_{\mu}(T,0)\Pi^\mu \phi({\bf x}_1) |0\rangle  \nonumber\\
&=& \frac{1}{{\rm
Det}[Q'_{\mu}({\bf x'}_2)-Q_{\mu}({\bf x}_1)]^{\Delta}} {\rm Det}[(i \tilde{a}_3
Q_{\mu}({\bf x}_2) + \tilde{a}_4 I)^{-1}]^{\Delta} \nonumber\\
&=& \frac{1}{{\rm Det}[(\tilde{a}_1 Q_{\mu}({\bf x}_2) -i \tilde{a}_2 I)-Q_{\mu}({\bf
x}_1)(i \tilde{a}_3 Q_{\mu}({\bf x}_2) + \tilde{a}_4 I)]^{\Delta}}.  \label{cexp2}
\end{eqnarray}
Note that the second term in the denominator vanishes when $x_{1
\mu} \to 0$. Further when $x_{2 \mu} \to \infty$, we find $C_1 \to
1/{\rm Det}[\tilde{a}_1 Q_{\mu_1}({\bf x}_2)]^\Delta$. Furthermore similar analysis shows
for multiple subsequent transformations given by
\begin{eqnarray}
U_{\mu_i}\Pi^\mu &=& \left( \begin{array}{cc} \tilde{a}_i & \tilde{b}_i \\ \tilde{c}_i & \tilde{d}_i
\end{array} \right), \quad \tilde{a}_i \tilde{d}_i-\tilde{b}_i \tilde{c}_i=1 \label{mult1}
\end{eqnarray}
$C_1 \to 1/{\rm Det}[(\prod_i \tilde{a}_i) Q_{\mu_i}({\bf x}_2)]^\Delta$. This allows us to
write the final expression for $F(T_0)$
\begin{eqnarray}
F(T_0) &=& \frac{1}{(\tilde{a}_0(T_0) \tilde{a}_3(T_0) \tilde{a}_1(T_0))^{\Delta}}
\label{fidexp2}
\end{eqnarray}
\begin{figure}[h]
\centering
\rotatebox{0}{\includegraphics*[width= 0.62 \linewidth]{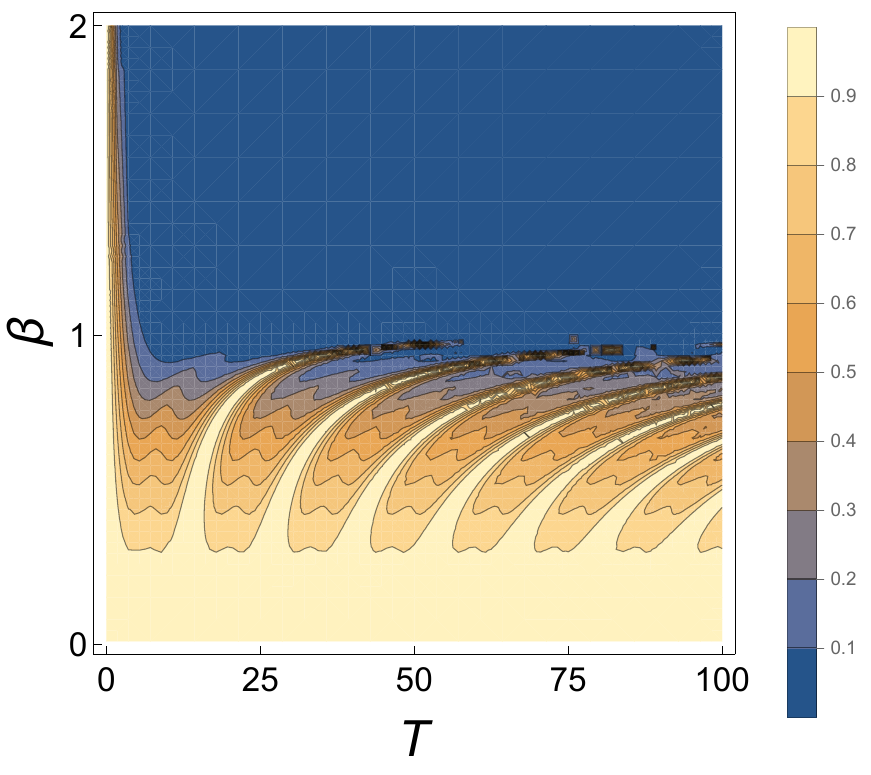}}
\caption{Plot of $|F(T)|$ as a function of $\beta$ and $T$ with $\Delta = 1$. For
$\beta<1$, $|F(T)|$ oscillates with $T$ characterizing the
non-heating phase and reaches $\simeq 1$ for $\nu= n \pi$ where $n$
is an integer. For $\beta>1$, $|F(T)|$ decays exponentially with $T$
which is a signature of the heating phase.\label{fig2}}
\end{figure}
%
To compute the $\tilde{a}_0$, $\tilde{a}_1$ and $\tilde{a}_3$, we use the $2 \times 2$
matrix representation of the operators $D$, $K_{\mu}$ and $P_{\mu}$
given in Eq.\ \ref{rep1}. Using this one can write
\begin{eqnarray}
U_{\mu}(\tau) \Pi^\mu &=& e^{-\tau \sqrt{1-\beta}^2(-n_z \sigma_z + n_x
\sigma_x)}, \quad n_z= \frac{1}{\sqrt{1-\beta^2}}, \quad n_x=
\frac{i\beta}{\sqrt{1-\beta^2}}  \label{rep2}
\end{eqnarray}
Defining $\nu = \sqrt{1-\beta^2} T/4$, we find, after analytically
continuing to real time $T=-iT_0$ {\color {black} and using $ {\tilde a}_{\mu}(T_0)= a_{\mu}(T)$}  
\begin{eqnarray}
a_0(T) &=& a_3(T) = (\cos \nu + i T(4\nu)^{-1} \sin \nu), \quad
a_1(T)=
(\cos 2\nu - i T (2\nu)^{-1} \sin 2\nu) \nonumber\\
F(T) &=& \frac{1}{[(\cos \nu + i T(4\nu)^{-1} \sin \nu)^2 (\cos 2\nu
- i T (2\nu)^{-1} \sin 2\nu)]^{\Delta}}  \quad  {\rm for} \, \,
\beta \ne1
\nonumber\\
&=& \left(\frac{1}{(1+iT/4)^2(1-iT/2)}\right)^\Delta \quad {\rm for}\, \,
\beta=1\label{ftfinal}
\end{eqnarray}

The behavior of each of the $a_i$ and hence $F(T)$ depends crucially on the value of the parameter $\beta$. For $\beta < 1$ these are oscillatory functions of the time $T$, while for $\beta > 1$ they decay exponentially for large $T$. These two behaviors are higher dimensional versions of the non-heating and heating phases. In between these phases there is a critical point  $\beta =1$ where we have a power law decay in time, $|F(T)| \sim 1/[(1+T^2/16)^2(1+T^2/4)]^{\Delta/2}$. This behavior is charted out in Fig \ref{fig2} which shows the behavior $|F(T)|$ for $\Delta=1$. In between, at $\beta=1$, we find a line which represents a critical line separating the
two phases.

Our analysis of $F(T)$ indicates special frequencies at which
$|F(T)|=1$ indicating perfect overlap of the driven state with the
initial states. These frequencies, which exist only in the
non-heating phase  can be read off from Eq.\ \ref{ftfinal} and are
given by $T=T_n^{\ast}(\beta)$ where
\begin{eqnarray}
T_n^{\ast}(\beta) &=& 4n \pi/\sqrt{1-\beta^2}, \label{spfreq1}
\end{eqnarray}
and $n$ is an integer. The existence of these frequencies can also
be seen in Fig.\ \ref{fig2}.

Thus we find that under a drive schematically represented by Fig.\
\ref{fig1}, $F$ exhibits perfect revival at special frequencies in
the non-heating phase and an exponential (power-law) decay with $T$
in the heating phase (on the critical line).

The different behaviors for different values of $\beta$ reflect the properties of the Hamiltonians used for time evolutions which have been discussed in the previous section. Each of these Hamiltonians represent a {\em different} $SU(1,1)$ subgroup of the conformal group. We chose the same $\beta$ for each of the three Hamiltonians, so the conjugacy classes of each of these $SU(1,1)$ are the same. The results can be trivially extended to the cases where the three $\beta$'s are different.

{ The above procedure can be also used to calculate transition amplitudes between a primary state and its descendants. In $1+1$ dimensions there are eficient algorithms to do this \cite{brehm}: it will be interesting to see if an analogous proecure can be developed in higher dimensions \footnote{We thank the referee for a question about this issue.}.}

\subsection{Unequal-time Correlators}
\label{corrsec}

In this section, we compute the unequal-time two-point correlation
function of the primary fields with conformal dimension $\Delta$ in
the presence of a drive. We consider the fields on a cylinder and
map them onto the plane using Eq.\ \ref{tran1}. The initial
coordinates on the plane corresponds to
\begin{eqnarray}
{\bf x_2}&=& (\tau_2,x_2,y_2,z_2)= (\cos \theta_2, \sin\theta_2 \cos
\phi_2, \sin \theta_2 \sin \phi_2 \cos \psi_2, \sin
\theta_2 \sin\phi_2 \sin \psi_2), \nonumber\\
{\bf x}_1 &=& (\tau_1,x_1,y_1,z_1) = (1,0,0,0) \label{cord1}
\end{eqnarray}
where we have taken initial time on the cylinder $w=0$ without loss
of generality. We have also chosen the coordinates $\theta_1=0$ for
simplicity. In what follows we shall consider the correlation
function
\begin{eqnarray}
C_{1}(T) &=& \langle 0 | U^{\dagger}(T,0) O({\bf x}_2) U(T,0) O({\bf
x}_1) |0\rangle  = \frac{J_2^{\Delta/4}}{({\rm Det}[Q({\bf x}'_2)
-Q({\bf x}_1)])^{\Delta}}  \label{cexp1}
\end{eqnarray}
where ${\bf x}'_2$ represents the transformed coordinate and $J_2$
denotes the Jacobian of the coordinate transformation.

\begin{figure}
\rotatebox{0}{\includegraphics*[width= 0.45 \linewidth]{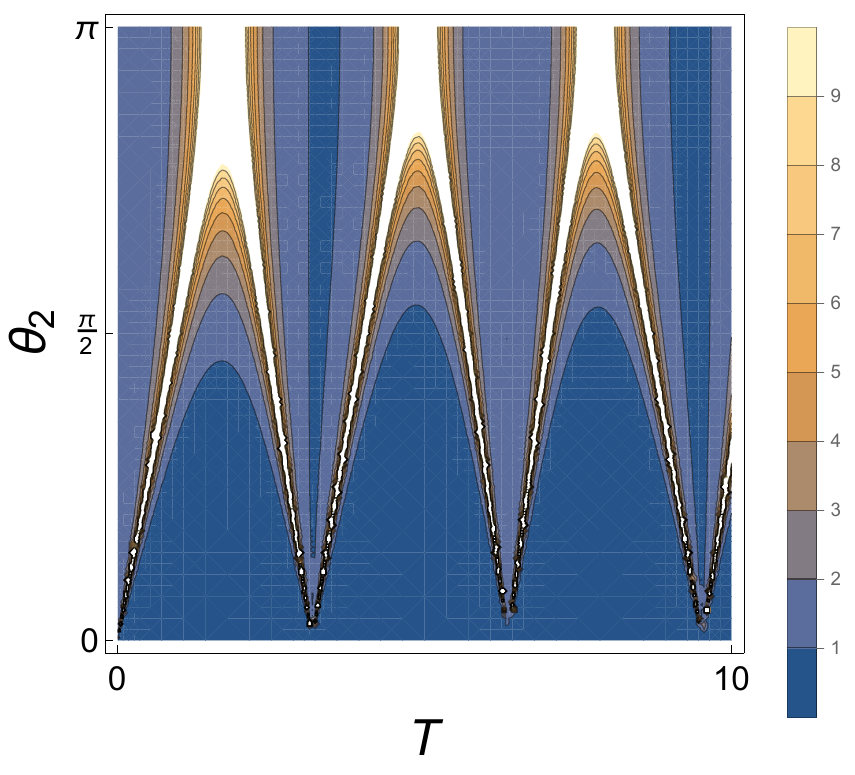}}
\rotatebox{0}{\includegraphics*[width= 0.45 \linewidth]{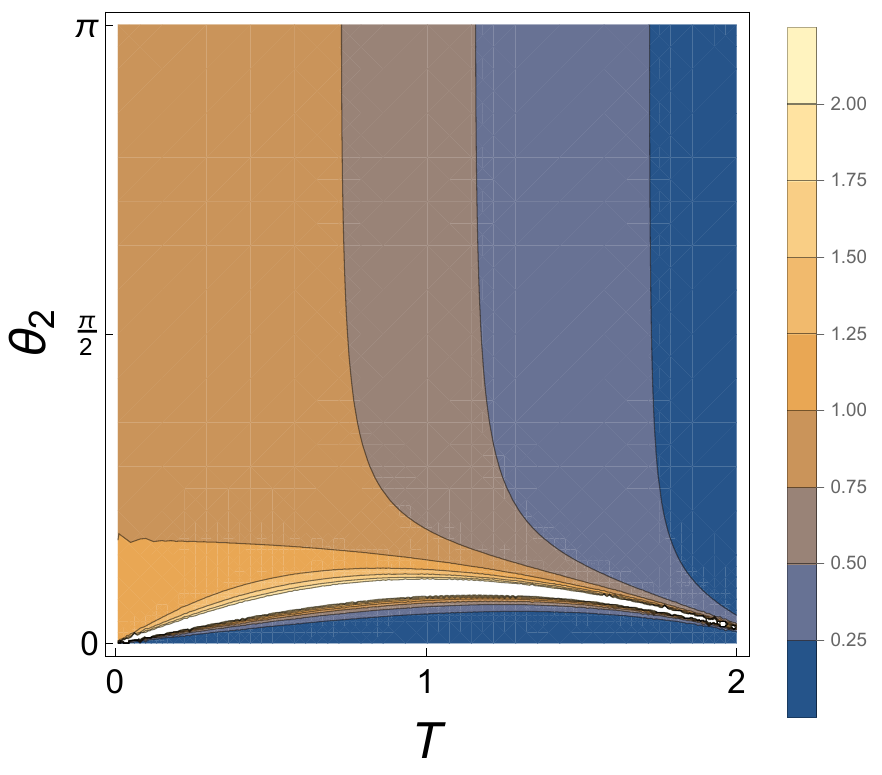}}
\caption{Plot of $|C_1(T)/C_1(0)|$ as a function of $\theta_2$ and
$T$ for $\beta=0.2$ (left panel) and $\beta_1=1.2$ (right panel)
corresponding to a square pulse protocol. See text for
details.\label{fig3}}
\end{figure}

%
%

To study the dynamics, we first consider a square pulse protocol given by
\bea
H & = & H_{(-)} = 2iD~~~~~~{\rm for}~~ t\le T/2 \nonumber \\
H & = & H_{(+)}= 2iD + i\beta \left(K_{0} +P_{0}\right) ~~~~~~{\rm for}~~ t>T/2.
\label{4-1}
\eea
We will calculate the correlator at time $T$.

The evolution operator for this is
\begin{align}
U(T,0) &= U_0(T,0) = e^{-T_0 \left(2 iD + i \beta \left(K_{0}+P_{0} \right) \right)/2} e^{-i D
T_0} = \left(\begin{array}{cc} \tilde{a} & \tilde{b} \\ \tilde{c} & \tilde{d} \end{array} \right)
\nonumber\\
a &= d^{\ast} = \left(\cosh \eta T/2 + \frac{i}{\eta} \sinh \eta T/2 \right)e^{iT/2}, \quad
b=c^{\ast}= e^{-iT/2} \frac{\beta}{\eta}\sinh \eta T/2,
\label{eq:abcd}\\
Q_{0}({\bf x}'_2) &=  (a Q_{0}({\bf x_2}) - i b I).(i c Q_{0}({\bf x}_2) + d
I)^{-1}, \,\,  J_2= |({\rm Det}[i c Q_{0}({\bf x}_2) + d
I])^{\Delta}|^{-4} \label{beq1}
\end{align}
where $\eta = \sqrt{\beta^2-1}$ and $I$ is the $2 \times 2$ identity
matrix. Here we have used Eqs.\ \ref{utrans1} and \ref{qtrans2}, { 
performed a Wick rotation $T=-i T_0$, and denoted $\tilde a(T_0),\, \tilde b(T_0), \,\tilde c(T_0)$, and $ \tilde d(T_0)$ 
after the rotation as $a(T),\,b(T),\,c(T),$ and $d(T)$ respectively.} Substituting Eqs.\ \ref{beq1} in
Eq.\ \ref{cexp1}, we find after transforming back to cylinder
coordinates \footnote{The Weyl factors cancel in the following expressions}
\begin{eqnarray}
C_{1}(T) &=& C_1(0) \left[ \frac{2(1-\cos \theta_2)}{(a - i c)^2 - (b - i
d)^2 - 2i(a - i c) (b - i d) \cos \theta_2}\right]^{\Delta}, \nonumber\\
C_1(0) &=& \frac{1}{2(1-\cos \theta_2)^{\Delta}} \label{correxp}
\end{eqnarray}
The dependence of $C_1(T)$ on both $T$ and $\beta$ can be obtained
by substituting the expressions of $a$, $b$, $c$ and $d$ from Eq.\
\ref{beq1}. We note that for $\beta>1$, the system is in the heating
phase and the correlation decays exponentially; in contrast, for
$\beta<1$, the dynamics is oscillatory. Also, we find that $C(T)$
depends only on $\theta_2$ and is independent of other coordinates.
This is a consequence of the choice of the drive protocol which
involves only $K_0$ and $P_0$. These features of the correlation
function are shown in Fig.\ \ref{fig3} where $|C_1(T)/C_1(0)|$ is
plotted for $\Delta = 1$ as a function of $\theta_2$ and $T$ for $\beta=0.2$ (left
panel) and $\beta=1.2$ (right panel).

We next consider a drive protocol involving two different sets
of generators given by
\begin{align}
H_{(-)} &= 2i D +i\beta(K_{(0)}+P_{(0)}), \quad  H_{(+)}= 2i D +i\beta(K_{(3)}+P_{(3)})
\nonumber\\
U(T_0,0) &= U_{(+)}(T_0,T_0/2) U_{(-)}(T_0/2,0), \quad U_{(\pm)}(\tau,0)=
e^{-\tau H_{(\pm)}/\hbar}  = \left(\begin{array}{cc} \tilde{a}_{\pm} & \tilde{b}_{\pm}
\\\tilde{c}_{\pm} & \tilde{d}_{\pm} \end{array} \right)  \label{hamsq}
\end{align}
To obtain the correlator corresponding to this protocol, we consider
the action of $U_-$ on $Q({\bf x}_2) \equiv Q_2 $ which is given by
\begin{eqnarray}
Q_{2}^{(1)} &=& (\tilde{a}_- Q_{2} -i \tilde{b}_-I).(i \tilde{c}_- Q_{2} + \tilde{d}_- I)^{-1} =
\left(\begin{array}{cc} \tau'-iz' & -i(x'-iy')
\\ -i(x'+ i y') & \tau'+i z' \end{array} \right) \nonumber\\
\tau' &=& \frac{ \tau(2\tilde{a}_- \tilde{d}_--1) -i (\tilde{a}_- \tilde{c}_- - \tilde{b}_- \tilde{d}_- r^2)}{\tilde{d}_-^2
+ 2 i \tilde{c}_- \tilde{d}_- \tau - \tilde{c}_-^2 r^2}, \quad x'_j = \frac{ x_{j}}{\tilde{d}_-^2 +
2 i \tilde{c}_- \tilde{d}_- \tau -\tilde{c}_-^2 r^2}  \label{firsttrans}
\end{eqnarray}
where $x_j = (x,y,z)$ for $j=(1,2,3)$. Next, we rewrite $Q_{2}^{(1)}$
as $Q_{2}^{'(1)} = z'I  -i \sigma_z \tau' - i \sigma_x x' - i\sigma_y
y'$ and perform the second transformation
\begin{eqnarray}
Q_{2}^{(2)} &=& (\tilde{a}_+ Q_{2}^{'(1)} -i \tilde{b}_+I).(i \tilde{c}_+ Q_{2}^{'(1)} + \tilde{d}_+
I)^{-1} = \left(\begin{array}{cc} z'' -i\tau '' & -i(x'' - iy'') \\ -i(x'' +  i y'') & z'' + i \tau'' \end{array} \right) \nonumber\\
z'' &=& \frac{ z'(2\tilde{a}_+ \tilde{d}_+ -1) -i (\tilde{a}_+ \tilde{c}_+ - \tilde{b}_+ \tilde{d}_+ r'^2)}{\tilde{d}_+^2 + 2
i \tilde{c}_+ \tilde{d}_+ z' - \tilde{c}_+^2 r'^2}, \quad x''_j = \frac{ x'_{j}}{\tilde{d}_+^2 + 2 i
\tilde{c}_+ \tilde{d}_+ z' - \tilde{c}_+^2 r'^2} \label{secondtrans}
\end{eqnarray}
where $x_j''=(\tau'',x'',y'')$ for $j=(1,2,3)$.

The rest of the calculation is cumbersome but straightforward. A
somewhat lengthy algebra yields
\begin{align}
C_2(T) &= C_1(0) \left( \frac{2(1-\cos \theta_2)}{\alpha + \beta_1 \cos
\theta_2 + \beta_2 \sin \theta_2 + \beta_3 \cos 2 \theta_2+ \gamma(\theta_2) \sin \theta_2 \sin \phi_2 \sin \psi_2}\right)^\Delta. \label{corr2}
\end{align}
Here $\alpha$, $\beta_i$ ($i=1,2,3$), and $\gamma$ are functions of the drive parameters through
functions $a$, $b$, $c$ and $d$ given by
\begin{eqnarray}
a &=& d^{\ast} = \left(\cosh \eta T/2 + \frac{i}{\eta} \sinh \eta T/2 \right), \quad
b=c = \frac{\beta}{\eta}\sinh \eta T/2,\label{corr3} \\
\gamma(\theta_2) &=& -4 i b (a - d) (b^2 - d^2 - 2 i b d \cos \theta_2), \quad \beta_2= b^2-a^2 \nonumber\\
\beta_3 &=& \frac{1}{2}\Big[8 a^3 d (1- a d)+ a^2 (-1 + 8 (b + 2 i d) d^2) - 8 i a b d (d + b (2 b d-i)) \nonumber\\
&& + b^2(1 + 8 d (i b - d(b^2 - d^2)))\Big]\nonumber\\
\beta_1 &=& 4 i \Big[2 a^4 b d - c^3 b(1 + 2d^2) +
   a (b^3 + i b^2 d (1 + 4 b^2) + 2 b^3 d^2 - i d^3) - b^3 d (1 + 4 d^2)\nonumber\\
   && +
   a^2 d (b - 2 b^3 - 4 i b^2 c4 + 2 i d^3) +
   b (2 b^4 d + i b d^2 + 2 d^5 - i b^3 (1 + 2 d^2) )\Big] \nonumber
\end{eqnarray}
where we have analytically continued to real time $T=-iT_0$. The function $\alpha$ is cumbersome
but can be expressed in terms of $a$, $b$, $c$ and $d$.

\begin{figure}[h]
\centering
\rotatebox{0}{\includegraphics*[width= 0.7 \linewidth]{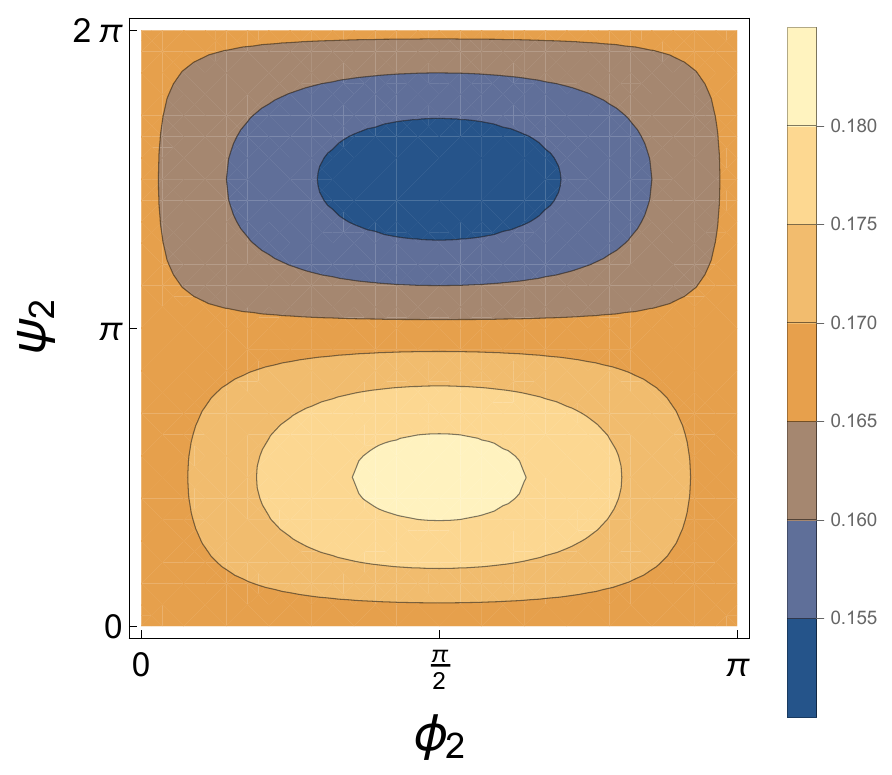}}
\caption{Plot of $|C_2(T)/C_2(0)|$ as a function of $\phi_2$ and
$\psi_2$ for $\beta=0.2$, $T=5$ and $\theta=\pi/4$. See text for
details. \label{fig4}}
\end{figure}

Eq.\ \ref{corr2} indicates a clear rotational symmetry breaking in
the correlation function. This is a consequence of application of
two different sets of generators $(K_{0},P_{0})$ and $(K_{3},P_{3})$ for
constructing $U(T,0)$. This dependence is shown for a fixed $T=1$,
$\theta_2=\pi/4$ and $\beta=0.2$ in Fig.\ \ref{fig4} where $|C_2(T)/C_2(0)|$
is plotted as a function of $\phi_2$ and $\psi_2$.

%

\subsection{Local probes of the driven state}
\label{local}
In this section we will probe the time-evolving state with local operators, i.e., we are interested in computing :
\begin{align}
\langle \Delta | U^\dagger_{\mu} (T,0)\Pi^\mu A(w^\alpha) U_{\mu} (T,0)\Pi^\mu |\Delta\rangle.
\label{3pt}
\end{align}
We will find here as in the $d=1$ case \cite{Fan:2019upv} that there is localization in the local observables.

\subsubsection*{Energy density}
\label{stressec}
Consider first the energy density, i.e. when the operator $A(w^\alpha)$ is chosen to be $T_{ww}(w^\alpha)$. Therefore we will need the transformation of the stress tensor. For $d=3$, the conformal dimension of the energy-momentum tensor is $4$, so that the Weyl factor is simply $e^{4w}$. Along with the factors coming from the transformation of a spin-2 tensor, the energy density translates to
\begin{align}
T_{ww}(w^\mu) &= e^{4 w} \frac{\partial x^\rho}{\partial w}  \frac{\partial x^\sigma}{\partial w} T_{\rho \sigma} (x)+\frac{3 a }{8\pi^2},
\end{align}
where the last term comes from the Weyl anomaly. Conjugation of the $R^{d+1}$ stress tensor $T_{\rho \sigma}(x)$ gets with the deformed evolution operators $U_\mu(T,0)\Pi^\mu$ leads to a conformal transformation. In $D$ flat spacetime for a spin-2 conformal primary like the stress-tensor the transformation rule is:
\begin{align}
U^\dagger T_{\mu\nu}(x) U &= T'_{\mu\nu}(x) = J^{\frac{D-2}{D}} \frac{\partial x'^\rho}{\partial x^\mu} \frac{\partial x'^\sigma}{\partial x^\nu} T_{\rho \sigma}(x'),
\end{align}
where $J$ denotes the Jacobian of the conformal transformation. Finally we shall be left to compute a plane three point function involving a stress-tensor and two primaries. This is given by \cite{Penedones}:
\begin{align}
\vev{O(x_1) O(x_2) T^{\mu \nu}(x_3)} &= - \frac{4 \Delta}{6\pi^2} \frac{H^{\mu\nu}(x_1,x_2,x_3)}{ |x_{12}|^{2\Delta-2} |x_{13}|^{2} |x_{23}|^{2}}, \\
\text{where, } H^{\mu \nu} &= V^\mu V^\nu - \frac{1}{4} V^2 \delta^{\mu\nu}, \,\,\, \text{with,  } V^\mu = \frac{x_{13}^\mu}{x_{13}^2} - \frac{x_{23}^\mu}{x_{23}^2} . \nonumber
\end{align}
Putting everything together, the time-dependent piece (i.e. modulo the anomalous piece) in the energy density of the driven state is:
\begin{align}
\vev{ \Delta | U^\dagger T_{ww}(w^\mu) U | \Delta }_{T} &= -2\frac{\Delta}{3\pi^2} \frac{e^{4w}J_2^{-1/2}}{|x'(w^\mu)|^2}\frac{ \partial x^\rho}{\partial w}\frac{ \partial x^\sigma}{\partial w}\frac{ \partial x'^\alpha}{\partial x^\rho} \frac{\partial x'^\beta}{\partial x^\sigma}  H_{\alpha \beta} ( x' ( w^\mu) ), \label{eq58} \\
\text{where, } H_{\rho \sigma}(x') &= \frac{ x'_\rho x'_\sigma - \frac{x'^2}{4} \delta_{\rho\sigma} }{ (x')^4 }. \nonumber
\end{align}
%
We consider a square pulse drive given by:
\bea
H & = & H_{(-)} = 2iD~~~~~~{\rm for}~~ t\le T/2 \nonumber \\
H & = & H_{(+)}= 2iD + i\beta \left(K_{(1)} +P_{(1)}\right) ~~~~~~{\rm for}~~ t>T/2.
\label{eq:x-pulse}
\eea
For this protocol the evolution operator is parametrized as : $ \left(\begin{array}{cc} a & b \\ c & d \end{array} \right)$ with $a,b,c,d$ as in Eq.\ \ref{eq:abcd}. Below in Fig.\ \ref{fig5} and Fig.\ \ref{fig6} we plot the absolute value of the normalized energy density
\begin{align}
E(\theta,\phi,\psi,T) = \frac{ \vev{ \Delta | U^\dagger T_{ww}(w^\mu) U | \Delta }_{T} }{\vev{ \Delta | U^\dagger T_{ww}(w^\mu) U | \Delta }_{T=0}}
\end{align} 
in various regimes.  Note that $\beta$ denotes the amplitude of the deformation and can be used to enter and leave the heating regime. When $\beta > 1$ and we are in the heating regime we find clear signatures of localization of the energy density in the angular directions, whereas the non-heating regime is characterized by oscillations. Furthermore the localization occurs in both the angular directions $\theta$ and $\phi$ as is clear from the density plot of Fig.\ \ref{fig7}, and is independent of the $\psi$ direction. {\color{black}Due to the latter feature we omit $\psi$ from $E(\theta,\phi,\psi,T) $. }

\begin{figure}
\rotatebox{0}{\includegraphics*[width= 0.42 \linewidth]{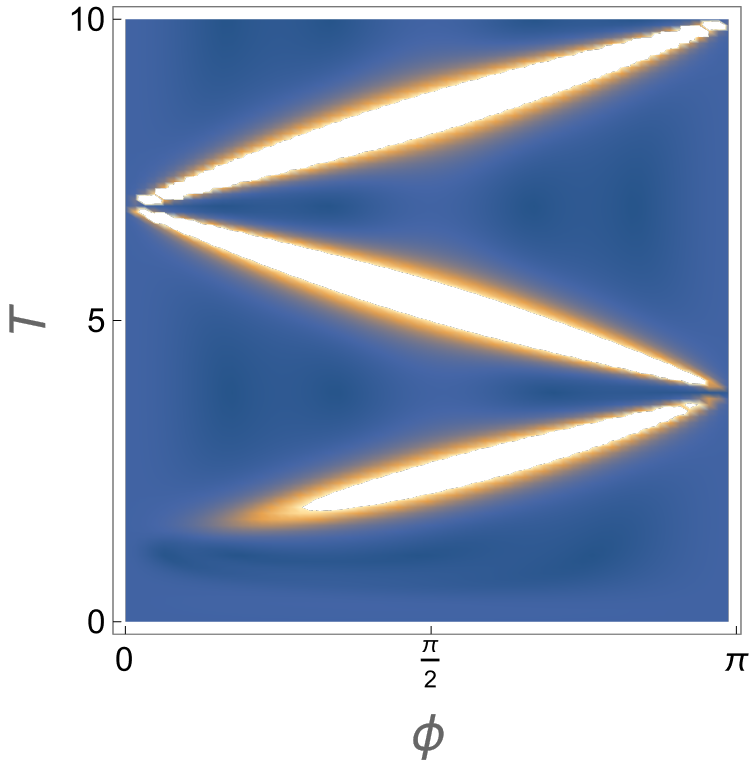}}
\rotatebox{0}{\includegraphics*[width= 0.04 \linewidth]{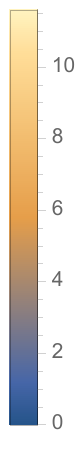}}
\rotatebox{0}{\includegraphics*[width= 0.42 \linewidth]{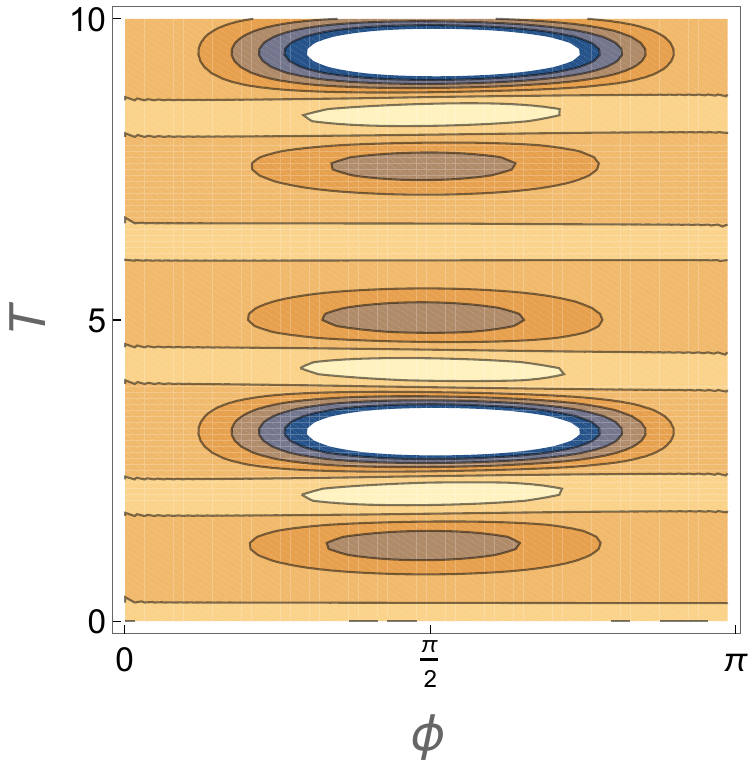}}
\rotatebox{0}{\includegraphics*[width= 0.04 \linewidth]{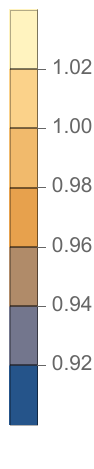}}
\caption{Density plots of normalized density $|E(\theta, \phi,T)|$ in the $\phi$, $T$ plane for $\theta=\pi/2$. Left Panel: In the heating phase with
$\beta=1.2$ we find progressive localization. Right panel: In the non-heating phase with
$\beta=0.09$ we find oscillations. } \label{fig5}
\end{figure}

\begin{figure}[h]
\rotatebox{0}{\includegraphics*[width= 0.8 \linewidth]{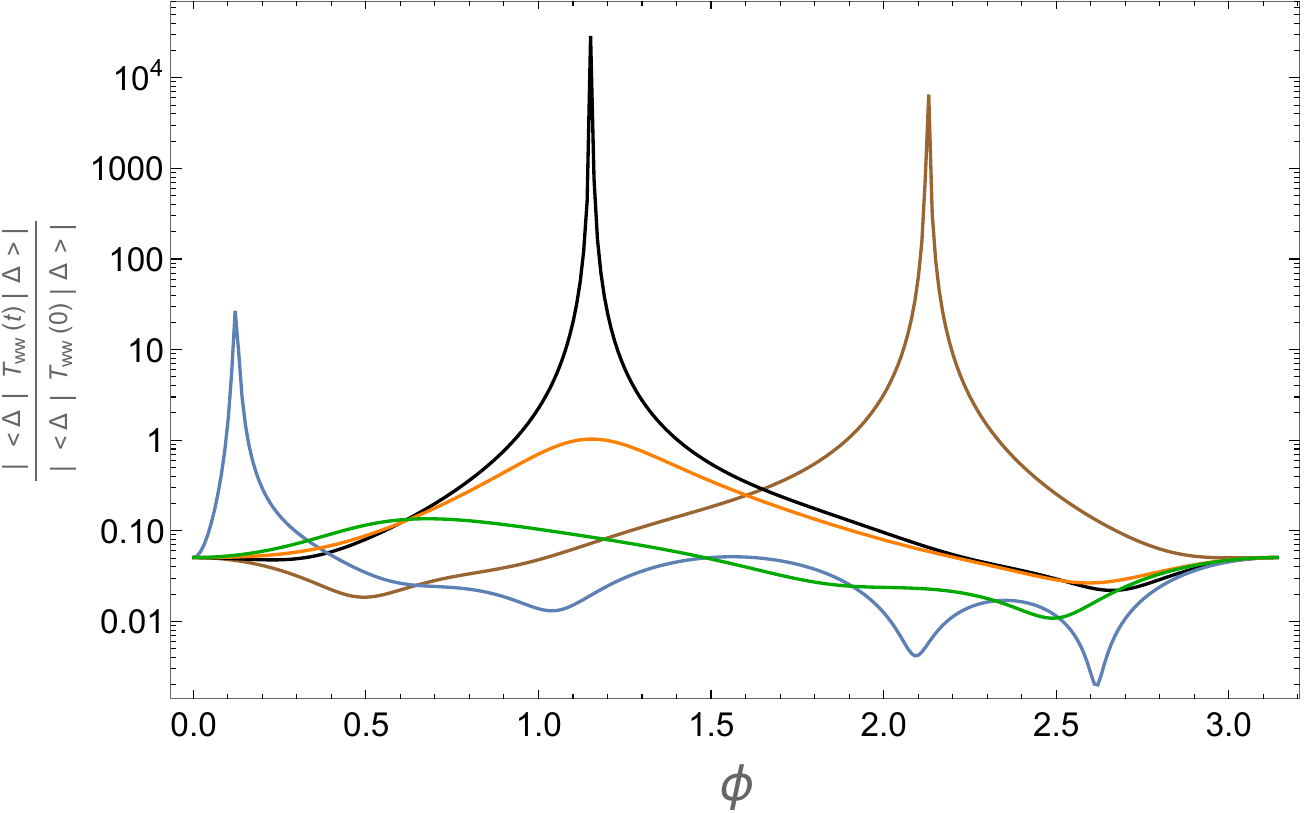}}
\caption{Plots of normalized density $|E(\theta, \phi,T)|$ as a function of $\phi$ for $ \theta=\pi/2$ in the heating phase, $\beta = 1.2$, and for
several representative values of $T$ : 1.5 (Green), 2 (Orange), 7 (Blue), 9 (Brown) and 12 (Black). \label{fig6}}
\end{figure}

\begin{figure}[h]
\begin{center}
\rotatebox{0}{\includegraphics*[width= 0.42 \linewidth]{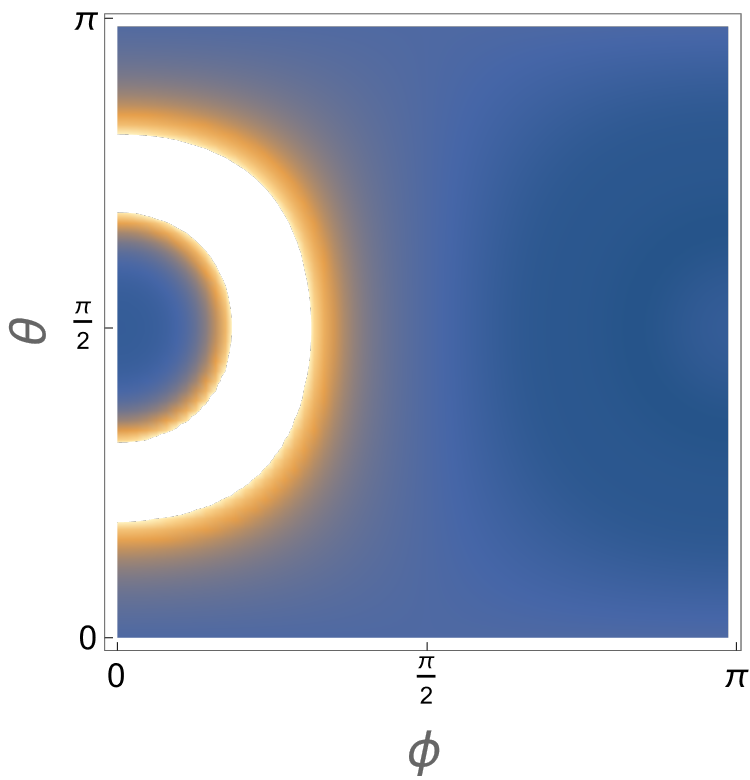}}
\rotatebox{0}{\includegraphics*[width= 0.04 \linewidth]{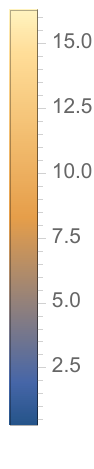}}
\caption{Density plots of normalized density $|E(\theta, \phi,T)|$ in the $\phi$, $\theta$ plane for $T=6$, in the heating phase,
$\beta=1.1$. } \label{fig7}
\end{center}
\end{figure}

The explicit formula for the energy density as obtained from the analytic expression Eq.\ \ref{eq58} turns out to be too complicated. Thus we turn our attention to a simpler observable, namely, that of a local primary probe in the driven state, which allows for an explanation of the angular localization.

\subsubsection*{Primaries}
\begin{figure}[h]
\centering
\rotatebox{0}{\includegraphics*[width= 0.52 \linewidth]{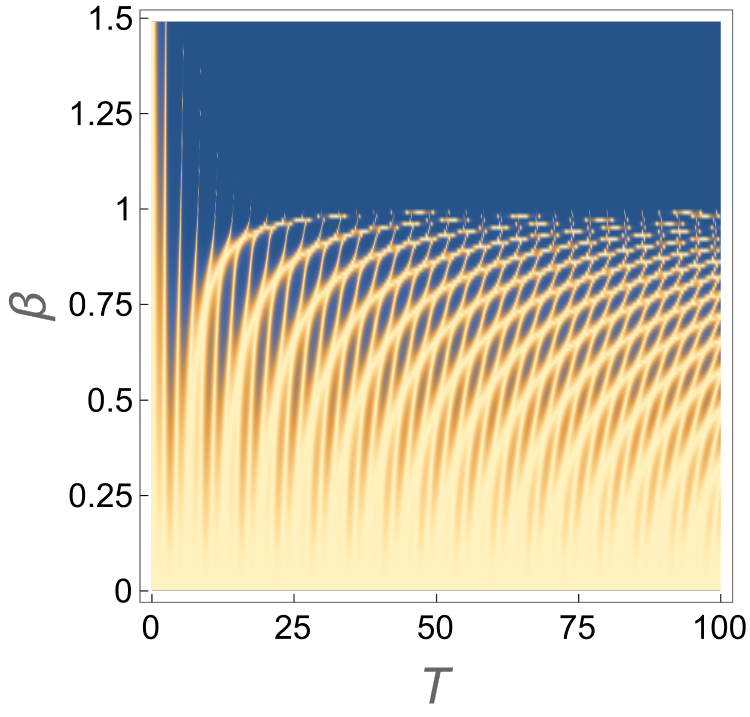}}
\rotatebox{0}{\includegraphics*[width= 0.05 \linewidth]{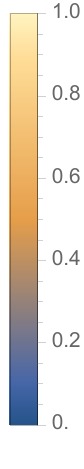}}
\caption{Plot of $|O(\theta=0, \phi,T)|_\Delta$ as a function of $\beta$ and $T$ with $\Delta_1 = 1$. For
$\beta<1$, we find oscillations with $T$ characterizing the
non-heating phase and for $\beta>1$, the one-point function decays exponentially with $T$
which is a signature of the heating phase.\label{fig7z}}
\end{figure}

Here we choose the probe to be another primary operator in place of $A(w^\alpha)$ in Eq.\ \ref{3pt}. The transformation rule of a primary is already given as in Eq.\ \ref{optran1}, using which along with the Weyl factor, Eq.\ \ref{tran1}, and the universal formula for three point primary correlator, we obtain:
\begin{align}
\langle \Delta | U^\dagger_{\mu} \Pi^\mu (T,0) O_{\Delta_1}(w^\alpha) U_{\mu}(T,0) \Pi^\mu  |\Delta\rangle &= e^{w \Delta_1} J_2^{\Delta_1/4} \frac{C_{\Delta \Delta \Delta_1}}{ | x' |^{\Delta_1} },
\end{align}
where $C_{\Delta \Delta \Delta_1}$ is the operator product expansion coefficient which is part of the CFT data. The above expression in the cylindrical coordinates, with initial time $w=0$, for a generic $SU(1,1)$ drive in the $x$ direction, takes the form :
\begin{align}
\vev{\Delta | O_{\Delta_1}(T) |\Delta} &= C_{\Delta \Delta\Delta_1}\frac{ [((1+a_2a_3)^2 - a_2^2 a_4^2 )^2 + 4 a_2^2 a_4^2 (1+a_2 a_3)^2 \cos^2 \phi \sin^2 \theta]^{-\Delta_1/4}}{a_4^{-4\Delta_1}[a_4^4(a_3^2 - a_4^2)^2 + 4 a_3^2 a_4^6 \sin^2 \theta \cos^2 \phi ]^{3\Delta_1/4}} \nonumber \\
&= C_{\Delta \Delta\Delta_1} O(\theta, \phi,T)_\Delta.
\end{align}
Once again there is independence in $\psi$. Notice that when either $\theta =0$ or $\phi = \pi/2$ the above becomes space independent. Once again, we consider the square-pulse protocol involving deformation in the $x$ direction, as in Eq.\ \ref{eq:x-pulse}. The parameter $\beta$ denotes the amplitude of the drive and can be used to enter and leave the heating regime. In Fig.\ref{fig7z} on the $\beta, T$ plane we obtain a density plot very similar to Fig.\ref{fig2}. In particular we notice, that oscillations die into exponential fall-offs as the $\beta =1$ line is crossed from below.
At $\beta=1$ the functional dependence of the absolute value squared of the one point function in the driven state is :
\begin{align}
&|O(\theta, \phi,T)|_\Delta^2 =  \bigg((8+4T^2 + T^4 + T^2(T^2-4)\cos 2T + 4T^3 \sin 2T)^2 - 4T^2 \cos^2 \phi ( T^2 ( 4 + T^2)^2 \nonumber \\
&+ ( 8 + 4T^2 + T^4)((T^2-4)\cos 2T + 4 T \sin 2T))\sin^2\theta + 4T^4(4+T^2)^2\cos^4\phi \sin^4 \theta\bigg)^{-\Delta_1}.
\end{align}
At large $T$, dropping the rapidly oscillatory pieces, we find:
\begin{align}
|O(\theta, \phi,T)|^2_\Delta &\sim \frac{1}{ T^{8\Delta_1} (1-2 \cos^2\phi \sin^2 \theta  )^{2\Delta_1 } } + {\cal O}(T^{-10\Delta_1} ).
\end{align}
It turns out that the higher order terms in the $1/T$ series (with rapidly oscillating terms dropped) contains higher order singularities of $1-2\cos^2 \phi \sin^2 \theta$ :
\begin{align}
|O(\theta, \phi,T)|^2_\Delta &= \sum_{n=1}^\infty  \frac{f_{n,\Delta_1}(\cos^2\phi \sin^2\theta)}{ T^{\Delta_1(6+2n)} (1- 2 \cos^2\phi \sin^2 \theta  )^{\tfrac{1}{2}(2n+1-(-1)^n)\Delta_1 } }.\label{eq63}
\end{align}
These singularities give rise to localization in the angular directions, and as we numerically investigate next, also persists in the heating, $\beta >1$ regime.

Below in Fig.\ref{fig8} we plot the normalized one point function amplitude in various regimes. We find clear signatures of localization of the amplitude in the angular directions, where as the non-heating regime is characterized by oscillations.
\begin{figure}[h]
\rotatebox{0}{\includegraphics*[width= 0.4 \linewidth]{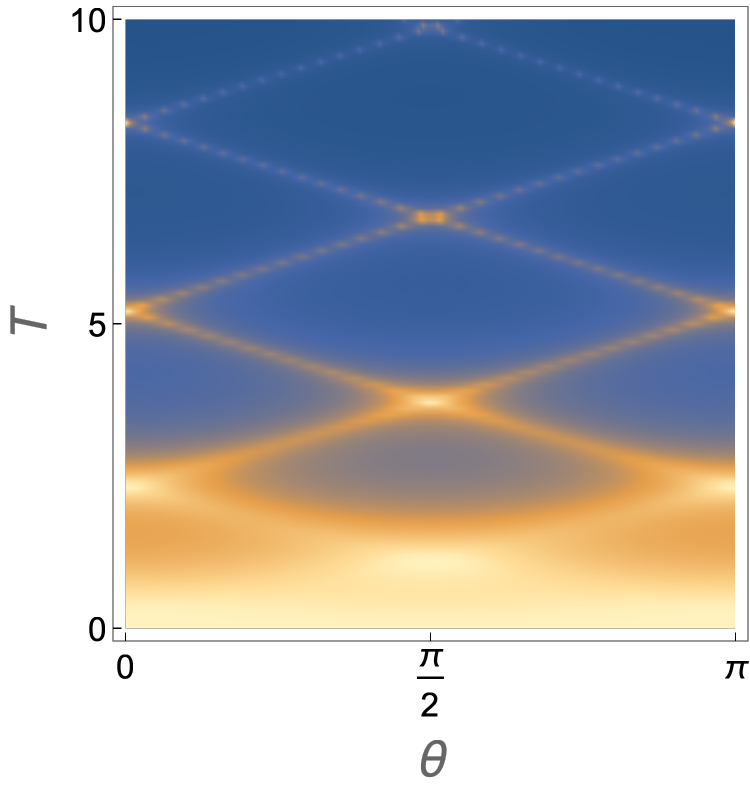}}
\rotatebox{0}{\includegraphics*[width= 0.05 \linewidth]{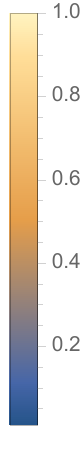}}
\rotatebox{0}{\includegraphics*[width= 0.4 \linewidth]{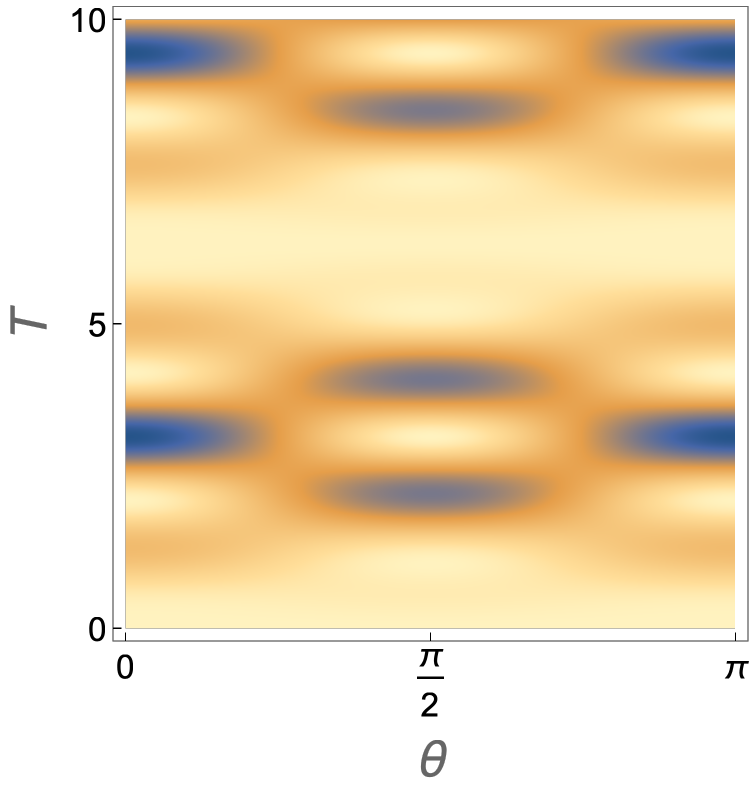}}
\rotatebox{0}{\includegraphics*[width= 0.06 \linewidth]{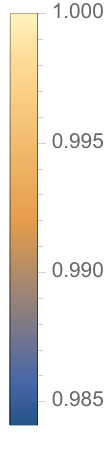}}
\caption{Density plots of normalized $|O(\theta, \phi,T)|_\Delta$ in the $\theta$, $T$ plane for $\phi=0$. Left Panel: In the heating phase with
$\beta=1.1$ we find progressive localization. Right panel: In the non-heating phase with
$\beta=0.1$ we find oscillations. The conformal dimension is taken to be $\Delta_1 = 0.4$. } \label{fig8}
\end{figure}
Localization occurs in both the $\phi$ as well as $\theta$ directions as is clear from the left panel of Fig.\ref{fig10}. In the right panel of Fig.\ref{fig10} we have plotted the contours of the function $\cos^2 \phi \sin^2 \theta$ since at least in the $\beta=1$ line we expect from Eq.\ \eqref{eq63} localization along the $\tfrac{1}{2}$ contour. The plots are indicative of the fact that similar singularities extend into the heating regime as well. It is very natural that a similar effect is responsible for the localization in the energy density when the primary gets replaced by the stress-tensor.
\begin{figure}[h]
\begin{center}
\rotatebox{0}{\includegraphics*[width= 0.42 \linewidth]{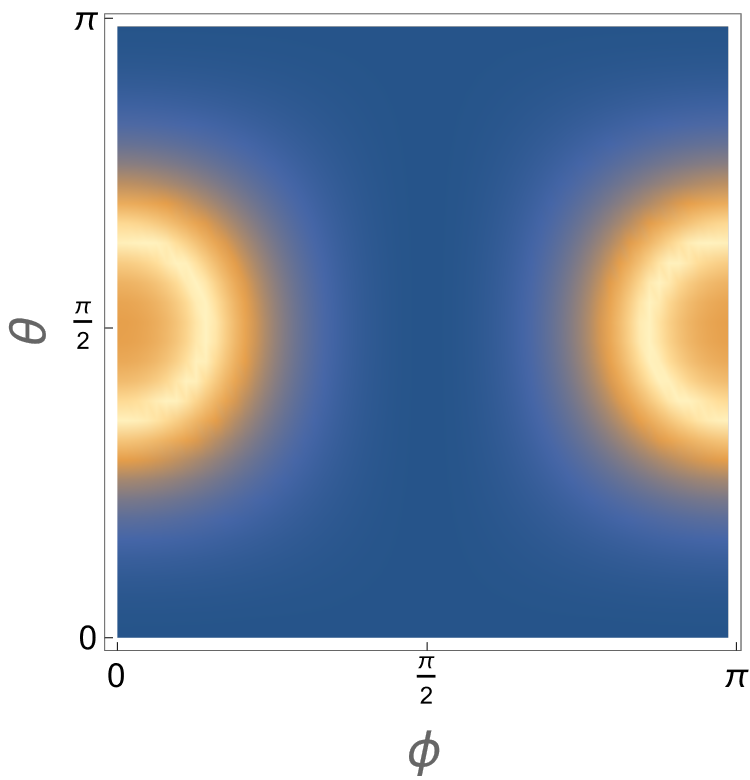}}
\rotatebox{0}{\includegraphics*[width= 0.05 \linewidth]{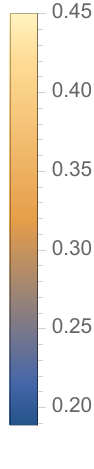}}
\rotatebox{0}{\includegraphics*[width= 0.42 \linewidth]{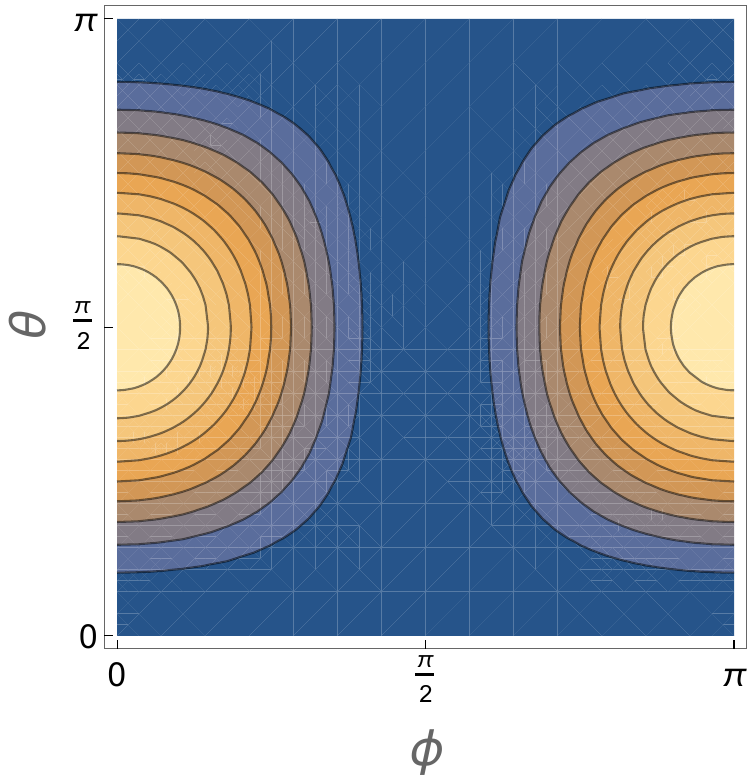}}
\rotatebox{0}{\includegraphics*[width= 0.045 \linewidth]{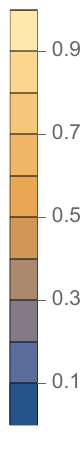}}
\caption{Left Panel : Localization from the density plots of normalized  $|O(\theta, \phi,T)|_{\Delta}$ in the $\phi$, $\theta$ plane for $T=3.3$, in the heating phase,
$\beta=1.1$. The conformal dimension is set to $\Delta_1 = 0.4$. Right Panel : Contour lines of $\cos^2 \phi \sin^2 \theta$. See text for details. } \label{fig10}
\end{center}
\end{figure}

\subsection{Floquet Hamiltonian for protocols involving a single  $SU(1,1)$ subalgebra and Dynamical Phases} 
\label{flha}

In this sub-section, we shall derive the Floquet Hamiltonian
for any periodic drive where the hamiltonians for each piece of a square pulse protocol all belong to the same $SU(1,1)$ subgroup of the conformal group generated by $D,K_\mu,P_\mu$. As explained earlier the real time evolution over a single cycle of time extent $T$ the evolution operator can be represented by a $SU(1,1)$ matrix
\bea
U(T,0) & = & \left (\begin{array}{cc} a & b \\ b^{\ast} & a^{\ast} \end{array} \right) \nonumber \\
|a|^2 - |b|^2 & = & 1
\label{umat1}
\eea
The parameters then determine the transformation of the quaternion as in (\ref{quatermobius}),
\ben
Q_\mu^\prime = (a Q_\mu -ib I)(ib^\star Q_\mu + a^\star I)^{-1}
\een
The Floquet Hamiltonian $H_F(T)$ is defined by (in $\hbar =1$ units)
\begin{eqnarray}
U(T,0) &=& \exp[-i H_F(T) T]. \label{umat2} 
\end{eqnarray} 
This should be a linear combination of the generators
\ben
H_F (T) = i\alpha_1 D +i\alpha_2 (K_\mu + P_\mu) +\alpha_3 (K_\mu - P_\mu)
\label{floquetform}
\een
Here $\alpha_1, \alpha_2, \alpha_3$ are real numbers. This is ensured by the hermiticity properties of the hermitian hamiltonians in each portion of the square pulse protocol. For s given protocol $H_F(T)$ can be obtained by using a Baker-Hausdorff-Campbell rule. However, since the result depends only on the algebra  one can instead use the representation  (\ref{rep1}) of the generators in terms of Pauli matrices $\sigma_i$. In this representation
\ben
H_F(T) \sim \sum_{i=x,y,z} \sigma_{i} \gamma_{i}
\label{hfgeneral}
\een
Note that it follows from the representation (\ref{rep1}) that one must have $\gamma_1,\gamma_2$ purely imaginary and $\gamma_3$ real. Comparing Eqs.\ \ref{umat1} and \ref{umat2} we can proceed to solve for $\gamma_i$. 

There are three classes, corresponding to the three conjugacy classes of the $SU(1,1)$ transformation. In the following we will denote
\ben
a_r= {\rm Re}[a], ~~~~~a_{I}= {\rm Im}[a], ~~~~~b_r= {\rm Re}[b],  ~~~~~b_{I}= {\rm Im}[b]
\label{arbr}
\een

\begin{enumerate}

\item $a_r^2 < 1$. This corresponds to $\sum_i \gamma_i^2 > 0$. In this case the transformation belongs to the elliptic conjugacy class. Then the Floquet hamiltonian is
\ben
H_F = \frac{\Lambda(T)}{\sqrt{1-a_r^2}} \left[  2i a_I D + ib_r (K_\mu + P_\mu) -b_I (P_\mu - K_\mu) \right]
\label{hfelliptic}
\een
where
\ben
T \Lambda_e(T) = \arccos (a_r)
\een

The evolution operator after $n$-cycles of the drive can be written as $ U(nT,0)=  \exp[-i n H_F(T) T]$. In matrix form
this can be written, for any integer $n$, as 
\begin{eqnarray} 
U(nT,0) &=&  \left (\begin{array}{cc} a_n & b_n \\ b_n^{\ast} & a_n^{\ast} \end{array} \right) \nonumber\\
a_n &=& \cos (n \Lambda_e T )+ i \frac{a_I}{\sin (\Lambda_e T)} \sin (n \Lambda_e T) \nonumber\\
b_n &=& \frac{b_r + i b_I}{\sin (\Lambda_e T)} \sin n \Lambda_e T \label{umatncyc}
\end{eqnarray} 

These coefficients are periodic in the number of cycles $n$ with a period $T_p$
\ben
T_p = \frac{2\pi}{\arccos (a_r)}
\een
so that the stroboscopic time evolution is periodic. The system is in a non-heating oscillatory phase.

\item $a_r^2 = 1$. We now have $\sum_i \gamma_i^2 = 0$ and the transformation is in the parabolic conjugacy class. In this case one has
\ben
H_F = 2i a_I D + ib_r (K_\mu + P_\mu) -b_I (P_\mu - K_\mu) 
\label{parabolic}
\een
At the end of $n$ cycles we have
\begin{eqnarray} 
U(nT,0) &=&  \left (\begin{array}{cc} a_n & b_n \\ b_n^{\ast} & a_n^{\ast} \end{array} \right) \nonumber\\
a_n &=& 1 + i n a_I  \nonumber\\
b_n &=& n(b_r + ib_I) \label{umatncyc-parabolic}
\end{eqnarray} 
As a function of $n$ these grow linearly in $n$, leading to a response which grows with Floquet time in a linear fashion.

\item $a_r^2 > 1$. In this case one has $\sum_i \gamma_i^2 < 0$ and the transformation is the hyperbolic conjugacy class. Now the Floquet hamiltonian is
\ben
H_F = \frac{\Lambda_h (T)}{\sqrt{a_r^2-1}} \left[  2i a_I D + ib_r (K_\mu + P_\mu) -b_I (P_\mu - K_\mu) \right]
\label{hfhyperbolic}
\een
where
\ben
T \Lambda_h (T) = {\rm {arccosh}} (a_r)
\een
The matrix which corresponds to time evolution by $n$ cycles is now given by
\begin{eqnarray} 
U(nT,0) &=&  \left (\begin{array}{cc} a_n & b_n \\ b_n^{\ast} & a_n^{\ast} \end{array} \right) \nonumber\\
a_n &=& \cosh n \Lambda_h T + i \frac{a_I}{\sinh (\Lambda_h T)} \sinh n \Lambda_h T \nonumber\\
b_n &=& \frac{b_r + i b_I}{\sinh (\Lambda_h T)} \sinh n \Lambda_h T \label{umatncyc-hyperbolic}
\end{eqnarray} 
For large stroboscopic times $n$ the response of the system now grow exponentially in the Floquet time $n$, characteristic of a heating phase.

\end{enumerate}

\begin{figure} 
\rotatebox{0}{\includegraphics*[width= 0.37 \linewidth]{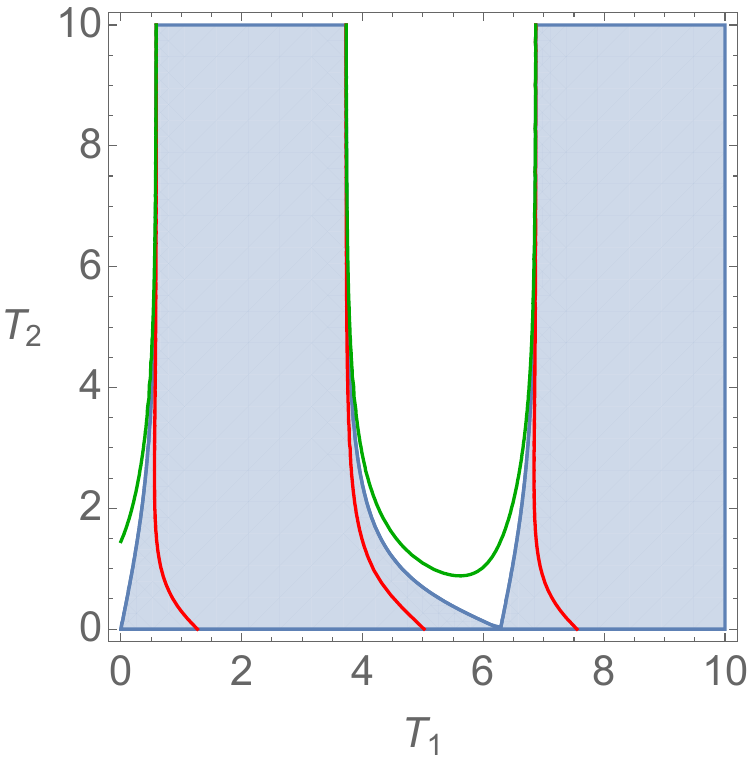}}
\rotatebox{0}{\includegraphics*[width= 0.37 \linewidth]{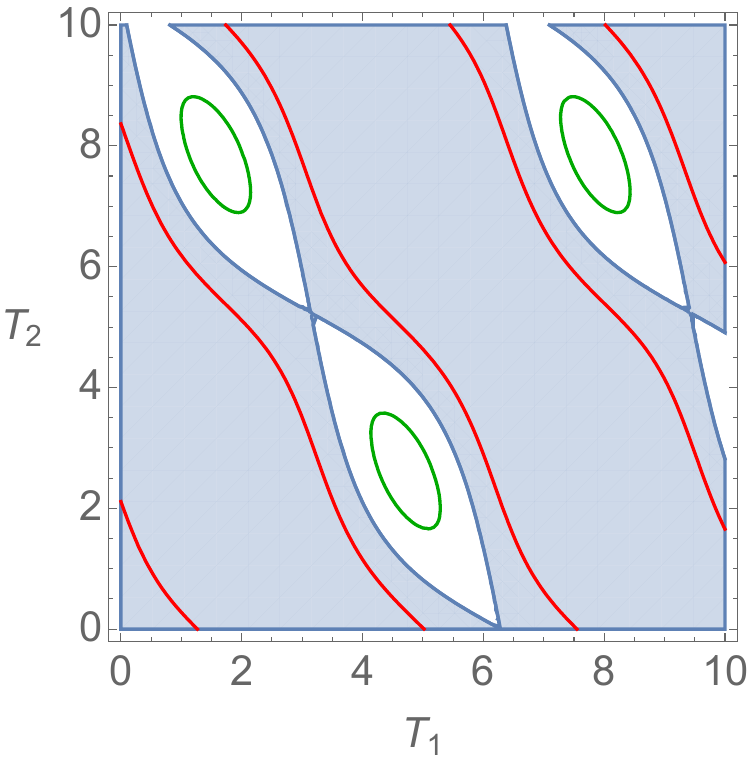}}
\rotatebox{0}{\includegraphics*[height= 0.35 \linewidth]{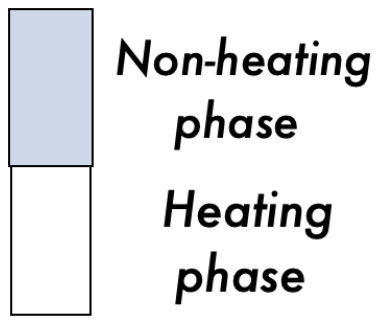}}
\caption{Contours of $({\rm Re}[a])^2$. Left Panel:  We use equation (\ref{ab-greater1}) with $\beta = 1.2$. The blue line is $({\rm Re}[a])^2 = 1$, the red line is $({\rm Re}[a])^2 = 0.3$ for and the green line has $({\rm Re}[a])^2 = 1.5$. Right panel: We use equation (\ref{ab-less1}) with $\beta = 0.8$. The blue line is $({\rm Re}[a])^2 = 1$, the red line is for $({\rm Re}[a])^2 = 0.3$ and the green line has $({\rm Re}[a])^2 = 1.5$. }
\label{fig:11}
\end{figure}

Since the phase in Floquet dynamics is determined by the conjugacy class of the corresponding conformal transformation it is possible to go between these phases by changing the time extents during which the different hamiltonians act within a single period. Consider for example a slight variation of the protocol (\ref{4-1}),
\bea
H & = & H_{(-)} = 2iD~~~~~~{\rm for}~~ t\le T_1\nonumber \\
H & = & H_{(+)}= 2iD + i\beta \left(K_{0} +P_{0}\right) ~~~~~~{\rm for}~~ T_1 \leq t \leq T_1 + T_2.
\label{4-1}
\eea
The period is now $T = T_1 + T_2$. The corresponding matrix elements of (\ref{umat1}) are given by, for $\beta > 1$,
\ben
a = e^{i T_1} \left( \cosh \eta T_2 + \frac{i}{\eta} \sinh \eta T_2 \right) ~~~~~~
b = e^{-i T_1} \frac{\beta}{\eta} \sinh \eta T_2, \quad \eta=\sqrt{\beta^2 - 1}
\label{ab-greater1}
\een
while for $\beta < 1$ we have
\ben
a = e^{i T_1} \left( \cos \nu T_2 + \frac{i}{\nu} \sin \nu T_2 \right) ~~~~~~
b = e^{-i T_1} \frac{\beta}{\nu} \sin \nu T_2, \quad \nu=\sqrt{1-\beta^2 }
\label{ab-less1}
\een
The conjugacy class is determined by the real part of $a$, one can have heating as well as a non-heating phases in Floquet dynamics for any value of $\beta$.
Left panel of Figure (\ref{fig:11}) shows a contour plot of $({\rm Re}[a])^2$ for $\beta = 1.2$, while right panel of Figure (\ref{fig:11}) shows a contour plot of ${\rm Re}[a]$ for $\beta = 0.8$.

These figures show how that can get dynamical phase transitions between heating and non-heating phases by changing $T_1$ and $T_2$ both for $\beta < 1$ and for $\beta > 1$.

Finally, the different dynamical phases can be also understood in terms of the structure of fixed points of the quaternionic Mobius transformations (restricted to a single $SU(1,1)$. This is discussed in Appendix (\ref{multiple}).

{ Note that we are using the word "heating" in a heuristically -  in the sense that the Floquet dynamics is exponentially damped as a function of the number of cycles.  Thermalization in conformal field theories are non-standard. In $1+1$ dimensions the presence of an infinite number of KdV charges implies that the system can evolve into a Generalized Gibbs Ensemble (see e.g. \cite{dymarsky2}). We are not aware of an analogous result in higher dimensions \footnote{We thank the referee for a comment about this point.}.}

\subsection{Floquet dynamics for protocols with different $SU(1,1)$ subgroups.}

The preceding results for Floquet dynamics were for the special case where the protocol is such that the transformation belongs to a single $SU(1,1)$ subgroup of the conformal group. When the protocol involves different $SU(1,1)$ subgroups in different time intervals, as in the calculation of fidelity above, the entries in the net quaternionic Mobius transformation (\ref{quatermobius} cannot be chosen to be complex numbers with any choice of the quaternion representation. Consider for example a two step square pulse protocol where we first use a hamitonian $H(\Pi^0)$ for $ 0 < t < T/2$ and a hamiltonian $H(\Pi^1)$ for $ T/2 < t < T$. In our calculations we dealt with this by switching the representation of quaternions, using $Q_0$ for the first interval and then collecting the transformed coordinates into a quaternion $Q_1$. However, as explained in Appendix B, we could have also used the quaternion representation $Q_1$ in the first step. This would lead to a transformed $Q_1^\prime$ given by (\ref{quatermobius}) where the entries are themselves quaternions. In the next time interval the transformation on $Q_1^\prime$ is simple since the entries are now numbers. More generally, the net result for composing two different quaternionic transformations take the form (\ref{composition}).

A derivation of the Floquet hamiltonian is likewise more involved. We leave this important aspect to a future publication.

\section{Discussions}
\label{conc}

In this paper we initiated a program of studying periodic quantum dynamics of driven deformed conformal field theories in arbitrary number of dimensions, inspired by recent work in $1+1$ dimensions. More concretely, we studied the properties of driven $3+1$ dimensional CFTs using square protocols which involve evolution by non-commuting Hamiltonians in successive time intervals after a single drive cycle.

There are several points worth emphasizing regarding this approach. First, the method of exactly computing the dynamics works in arbitrary number of dimensions. We performed explicit calculations in $3+1$ dimensions in this work. But our use of quaternion formalism provides a simple way to perform these calculations all {\color{black}$d\le 3$}. Secondly, our method can be straightforwardly generalized to other drive protocols including continuous ones; we leave this as a subject of future work. For a class of drive protocols which uses same $\mu$ but different $\beta$ for each pulse within a drive cycle, we have shown how to obtain the necessary transformations for arbitrary number of cycles. This is similar to $d=1$, and we can have dynamic control
over the transition from the heating to the non-heating phase similar to that found in $d=1$ driven CFTs \cite{Wen:2018agb, Wen:2020wee, Das:2021gts}.
Explicit results for multiple cycles of this type will appear in a future paper.
For pulses with different $\mu$, the situation is less clear
and this issue is left as a subject of future study.

While the power of conformal symmetry allows us to calculate observables and the details associated with the phase structure, perhaps it is useful to ponder over a simpler physical intuition of the underlying physics. In this regard, it is useful to consider {\it e.g.}~free theories in $(1+1)$-dimensions. For free Bosons, the stress-tensor $T(w) = (-1/2) :\partial\phi(w) \partial\phi(w):$ and for free Fermions it is given by $T(w)= (1/2) :\psi(w) \partial\psi(w):$ and subsequently the Virasoro modes are extracted from the stress-tensor: $L_n = (1/(2\pi i))\oint w^{n+1} dw T(w)$. The standard radial quantization in CFT chooses a Hamiltonian $H \sim L_0$, which corresponds to a stress-tensors of the form: $ \partial\phi(w) \partial\phi(w)$ for Bosons and $ \psi(w) \partial\psi(w)$ for Fermions. Choosing a different Hamiltonian involving $\{L_m, L_{-m}\}$-modes ({\it i.e.}~a different quantization) yields a non-trivial $w$-factor in front of the derivative terms in the stress-tensor and therefore will correspond to a non-trivial red-shift physics for the system. Consequently, thermal-phases may appear as a result of this.\footnote{Qualitatively, this is similar to the Rindler-physics, although it appears more subtle and layered.} Note, however, that this is too crude to pass as an argument since we have not factored in the inequivalent conjugacy classes and we have also been imprecise about where the ``red-shift" matters, {\it i.e.}~near the origin of the complex plane or near infinity. Nonetheless, it raises an intriguing possibility of a more precise physical picture along these lines and we hope to address this in future.

For holographic CFT's, it would be insightful to get a dual gravitational description of the dynamics. This can be done from two different perspectives: One that is based on the explicit time-dependent Hamiltonian and the other that is based on the Floquet Hamiltonian. While the former yields a dynamical geometry, the latter is capable of describing static ``patches" of the geometry which correspond to the different phases (heating and non-heating) and the phase boundary. For the $1+1$ dimensional case, the dynamical scenario has been explored in {\color{black} \cite{Goto:2021sqx, Goto:2023wai, Kudler-Flam:2023ahk}}, while the Floquet-Hamiltonian based approach is discussed in \cite{Das:2022pez}. In \cite{Goto:2021sqx}, a non-trivial dynamical horizon structure has been observed, whereas the different geometric patches of \cite{Das:2022pez} were obtained by solving integral curves that are generated by the (appropriate combination of) bulk Killing vector that is dual to the CFT Floquet Hamiltonian. Interestingly, the integral curve approach is also explored in a completely different context of ``bulk reconstruction" in which a local bulk observer in an AdS-geometry is described in terms of the CFT Hamiltonian, see {\it e.g.}~\cite{Jafferis:2020ora, Bahiru:2022oas}, .

The higher dimensional generalization of the bulk geometric description is potentially very interesting. First, it is rather non-trivial to obtain horizons with structures in higher dimensional black holes. Secondly, it is also technically involved to obtain dynamical horizons, outside Vaidya-type geometries. Moreover, from the integral curve perspective, a local bulk observer seems readily describable in terms of the CFT Floquet Hamiltonian using the bulk Killing vectors that describe the SL$(2,C)$ sub-algebra of the conformal algebra, see for example appendix \ref{EConf_d} for explicit expressions of the corresponding generators. We are currently exploring this aspect in detail.

Let us now offer some comments that are not necessarily limited to CFTs. It is evident that the crucial ingredients of our results are the following: (i) The existence of an SL$(2,C)$ algebra as a sub-algebra of the symmetry group of the system and (ii) a ``quasi-primary" representation of the fields under this SL$(2,C)$.\footnote{A similar statement holds for an SL$(2,R)$ algebra as well.} Interestingly, scattering matrix elements in $(3+1)$ and $(2+1)$-dimensions can be recast into correlation functions of quasi-primary operators at null infinity known as the celestial sphere, see {\it e.g.}~\cite{Pasterski:2016qvg, Pasterski:2017kqt, Crawley:2021ivb} for explicit details on this. These quasi-primaries form the continuous principal series, $\Delta = \frac{d-1}{2} + i s$, in $(d+1)$-dimensions and $s$ is real-valued; we have collected some relevant and explicit formulae in appendix \ref{CS_CFT} for a more direct comparison. Even though the ``CFT-spectrum" is continuous and complex-valued, the corresponding correlators are real-valued since it involves both $\Delta$ and $\bar{\Delta}$. Thus, the presence of a heating and a non-heating phase along with a phase boundary is expected in this case as well. Note, however, that this entails a rather non-trivial quantization of the system: the Minkowski coordinates, $X^\mu$, are first mapped to a stereographic coordinates on the null sphere, $w$; subsequently, the $w$-plane is mapped to the cylinder by $\zeta = {\rm exp}(2\pi w /L)$. Identifying $\zeta = T + i x$ and analytically continuing $T \to i T_L$ yields the Lorentzian system (with Lorentzian time $T_L$) for which the phases are expected to exist for different conjugacy classes of the SL$(2,C)$-transformations. It will be very interesting to explore this aspect as well as its connection with scattering matrix physics in detail, which we leave for future.

\section{Acknowledgements}

We would like to thank Bobby Ezhuthachan, Akavoor Manu, Masahiro Nozaki and Koushik Ray for discussions.
The work of S.R.D. is supported by a National Science Foundation grant NSF-PHY/211673
and Jack and Linda Gill Chair Professorship. S.R.D. would like to thank Yukawa Institute
of Theoretical Physics and Isaac Newton Institute for Mathematical Physics for hospitality
during the completion of this work. KS thanks DST for support through SERB project JCB/2021/000030. AK is partially supported by CEFIPRA $6304-3$, DAE-BRNS $58/14/12/2021$-BRNS and CRG$/2021/004539$ of Govt.~of India. D.D. acknowledges support by the Max Planck Partner Group grant MAXPLA/PHY/2018577 and from MATRICS grant {{SERB/PHY/2020334}}.

\appendix

\section{Transformation of the generators}
\label{appa}

In this section, we use infinitesimal coordinate transformations
to obtain expressions of the generators in the deformed frame as eluded to
in the main text. To this end, we recollect from the main text that
the transformation of the generators $D$, $K_0$ and $P_0$ can be easily obtained as follows
\begin{eqnarray}
U^{-1} \sigma_z U &=& \tau_z= \cosh \theta \sigma_z + \sinh \theta i
\sigma_x \nonumber\\
U^{-1} \sigma_x U &=& \tau_x = \cosh \theta \sigma_x - i \sigma_z
\sinh \theta, \quad  \tau_y = \sigma_y \label{utrans13}
\end{eqnarray}
where $U= \exp[-i \theta (K_0-P_0)/2]$ as defined in the main text and
the last relation follows from the fact that $U$ commutes with
$\sigma_y$. Eq.\ \ref{utrans13} therefore species the relations
between the generators $D$, $K_t$ and $P_t$ and their deformed
versions $D'$, $K'_0$ and $P'_0$. These are given by
\begin{eqnarray}
D' &=& D \cosh \theta -\frac{1}{2} (K_0 +P_0) \sinh \theta
\nonumber\\
K'_0 [P'_0] &=& \frac{1}{2}(1+\cosh \theta) K_0 [P_0] - \frac{1}{2}
(1-\cosh \theta) P_0 [K_0] - D \sinh \theta  \label{genrel1}
\end{eqnarray}
where we have used Eq.\ \ref{rep1} to obtain the generators from the
Pauli matrices.

To construct rest of the transformed generators, we need to use
their coordinate representation. For this, it is convenient to first
consider an infinitesimal transformation $U_{\rm inf}= 1 - i\theta
(K_0 -P_0)/2$. The transformation of coordinates $(\tau,x,y,z)\equiv
(\tau, x_j)$  under the action of $U_{\rm inf}$ can be computed and
yields
\begin{eqnarray}
\tau' &=& U \tau = \tau -\frac{\theta}{2}(1+r^2 - 2\tau^2),
\quad x'_j = U  x_j = x_j(1 + \theta \tau)  \label{ctrans1}
\end{eqnarray}
The reverse transformation expressing $(\tau, x_j)$ in terms of
$(\tau' , x'_j)$ can be obtained from Eq.\ \ref{ctrans1} and yields
\begin{eqnarray}
\tau &=&  \tau' +\frac{\theta}{2}(1+(r')^2 - 2\tau'^2), \quad x_j =
x'_j(1 - \theta \tau')  \label{ctrans2}
\end{eqnarray}

We now use the transformed coordinates to write the expressions of
the generators. For example, one finds
\begin{eqnarray}
P'_0 &=& -i \partial_{\tau'} = -i \partial_{\tau} +i \theta( \tau
\partial_{\tau} +\sum_j x_j \partial_j ) = P_0 - \theta D
\label{ptgen}
\end{eqnarray}
where we have used Eq.\ \ref{ctrans2}. Note that this coincides with
the infinitesimal version of Eq.\ \ref{genrel1} as expected; this
can also be explicitly checked for $D$ and $K_0$.

Next we find the other generators where the use of the transformed
coordinates provides us the infinitesimal version of the generators
in the deformed frame. For example, one finds
\begin{eqnarray}
P'_{j} &=& - i \partial_{x'_j} = - i (\partial_{x_j} + \theta (x_j
\partial_{\tau} - \tau \partial_{x_j}) ) = P_{j} + \theta L_{0j} \label{pxgen}  \\
K'_{j} &=& -i (x'_j (\tau' \partial_{\tau'} + \sum_j x'_j
\partial_{x'_j}) -(r')^2 \partial_{x'_j}) \nonumber\\
&=& -i (x_j (\tau \partial_{\tau} + \sum_j x_j
\partial_{x_j}) - r^2 \partial_{x_j}) -i \theta (\tau \partial_{x_j}
- x_j \partial_{\tau}) = K_{j} - \theta L_{0j} \label{kxgen}
\end{eqnarray}
where $L_{\mu \nu} = i (x_{\mu} \partial_{x_{\nu}} - x_{\nu}
\partial_{x_{\mu}})$ are the angular momentum generators. A simlar
analysis yields
\begin{eqnarray}
L'_{0j} &=&  L_{0j} +\frac{\theta}{2} (P_{j} - K_{j}), \quad L'_{ij}
=L_{ij}.  \label{amgen}
\end{eqnarray}
It can be easily checked that the generators obtained in Eqs.\
\ref{pxgen}, \ref{kxgen}, and \ref{amgen}, together with the
infinitesimal version of $D$, $K_0$ and $P_0$ obtained from Eq.\
\ref{genrel1}, satisfy the conformal algebra to ${\rm O}(\theta)$.

Next, we look for the transformed generators for finite conformal
transformation. Instead of a direct calculation via coordinate
transformations which is cumbersome, we use the results obtained for
infinitesimal transformations and the criteria that these generators
must satisfy the conformal algebra. It turns out that these two
criteria can uniquely specify the deformed generators. To see this,
we first consider $P_{j}$. Eq.\ \ref{pxgen} dictates the
infinitesimal version of $P'_j$; taking cue from this and Eq.\
\ref{genrel1}, we write
\begin{eqnarray}
P'_{j} &=& \frac{1}{2}(1+\cosh \theta) P_j + \frac{1}{2} (1-\cosh
\theta) K_j + L_{0j} \sinh \theta \label{pjgen}
\end{eqnarray}
The first check to ensure that this is the correct form is to verify
the commutator $[D', P_j]= i P'_j$. Written explicitly, using
standard commutation relation of conformal generators, this yields
\begin{eqnarray}
[D', P'_j] &=& i \Big( \frac{\cosh \theta(1+\cosh \theta)}{2} P_j -
\frac{\cosh \theta(1-\cosh \theta)}{2} K_j -\frac{1}{2} \sinh^2
\theta (P_j+K_j) \nonumber\\
&& + \sinh \theta L_{0j} \Big) = i P'_j \label{checkcomm}
\end{eqnarray}
Next, we need to check the commutation relation between $P'_j$ and
$K'_j$. For this one needs to know $K'_j$. Using Eq.\ \ref{kxgen}
and taking cue from the form of $P'_j$, we write
\begin{eqnarray}
K'_{j} &=& \frac{1}{2}(1+\cosh \theta) K_j + \frac{1}{2} (1-\cosh
\theta) P_j - L_{0j} \sinh \theta \label{pjgen}
\end{eqnarray}
This form ensures $[D', K'_j]= - i K'_j$ and this can be checked in
a similar manner. The commutation of $K'_j$ and $P'_j$ can be
computed to be
\begin{eqnarray}
[K'_j, P'_j] &=& 2i \left( \frac{(1+\cosh \theta)^2 - (1-\cosh
\theta)^2}{4} D - \sinh \theta (K_0 +P_0) \right) = 2 i D'
\label{checkkpcomm}
\end{eqnarray}
The commutation of $P'_0$ and $K'_j$ can also be checked and yields
\begin{eqnarray}
[K'_j, P'_0] &=& 2i \left( \frac{(1+\cosh \theta)^2 -(1-\cosh
\theta)^2}{4} L_{0j} + \frac{1}{2} \sinh \theta D (P_j-K_j) \right)
\nonumber\\
&=& 2i \left( \cosh \theta \, L_{0j} + \sinh \theta (P_j-K_j)/2
\right) = 2 i L'_{0j}  \label{ljgen}
\end{eqnarray}
Note that the form of $L'_{0j}= \cosh \theta \, L_{0j} + \sinh
\theta (P_j-K_j)/2$ so obtained is consistent with Eq.\ \ref{amgen}.
Further it can be checked that $[D', L'_{0j}]=0$. The other
commutation relations involving $L'_{ij}$ and $L'_{0j}$ also holds
provided we set $L'_{ij} = L_{ij}$. Thus the final expressions of
the deformed generators, used in the main text, are given by
\begin{eqnarray}
D' &=& D \cosh \theta -\frac{1}{2} (K_0 +P_0) \sinh \theta
\nonumber\\
K'_0 &=& \frac{1}{2}(1+\cosh \theta) K_0  - \frac{1}{2}
(1-\cosh \theta) P_0 - D \sinh \theta \nonumber\\
P'_0 &=& \frac{1}{2}(1+\cosh \theta) P_0  - \frac{1}{2} (1-\cosh
\theta) K_0 - D \sinh \theta \nonumber\\
K'_{j} &=& \frac{1}{2}(1+\cosh \theta) K_j + \frac{1}{2} (1-\cosh
\theta) P_j - L_{0j} \sinh \theta \nonumber\\
P'_{j} &=& \frac{1}{2}(1+\cosh \theta) P_j + \frac{1}{2} (1-\cosh
\theta) K_j + L_{0j} \sinh \theta  \nonumber\\
L'_{0j} &=& \cosh \theta \, L_{0j} + \frac{1}{2} \sinh \theta (P_j-K_j), \quad
L'_{ij}= L_{ij}  \label{allgen}
\end{eqnarray}
Thus Eq.\ \ref{allgen} provides a complete description of the
deformed CFT in terms of the new generators.

{

\section{Coordinate transformation involving different subgroups} 
\label{appb} 

In this subsection, we provide an alternative formulation for coordinate transformation involving different 
${\rm SU}(1,1)$ subgroups. To this end we define the initial quaternion matrix to be 
\begin{eqnarray}
Q &=& x_{\mu} I - i (\sigma_i x_{\nu} + \sigma_j x_{\lambda} + \sigma_k x_{\delta})\label{qinit1}
\end{eqnarray}
where $\sigma_{i,j,k}$ denote Pauli matrices with $\sigma_j \sigma_k = -i \sigma_i$, $I$ is the $2 \times 2$ identity matrix, and $\mu \ne \nu,\lambda,\delta$. 

Next we consider a transformation involving $D$, $K_{\nu}$, and $P_{\nu}$, where $\nu \ne \mu$. Let us consider a evolution operator 
$U(T,0)$ constructed using $D$, $K_{\nu}$ and $P_{\nu}$ given by
\begin{eqnarray}
U_{\nu} &=& \left( \begin{array}{cc} \tilde{a}_{\nu} & \tilde{b}_{\nu} \\ \tilde{c}_{\nu} & \tilde{d}_{\nu} \end{array} \right) \label{trans1-b}
\end{eqnarray} 
where $ \tilde{a}_{\nu} \tilde{d}_{\nu} - \tilde{b}_{\nu} \tilde{c}_{\nu} =1$. We consider a quaternion matrix $Q'$ which can be written as 
\begin{eqnarray} 
Q' &=&   x_{\nu} I - i (\sigma_{i} x_{\mu} + \sigma_j x_{\delta} + \sigma_k x_{\lambda}).\label{qinit2} 
\end{eqnarray} 
In this frame, the transformation of coordinates is known to be 
\begin{eqnarray} 
Q_{\rm new} &=& ( \tilde{a}_{\nu} Q' - i \tilde{b}_{\nu} I).(i \tilde{c}_{\nu} Q' + \tilde{d}_{\nu} I)^{-1} \label{qtran1} 
\end{eqnarray}
To obtain from $Q_{\rm new}$ starting from Q (Eq.\ \ref{qinit1}), we therefore need to transform $Q$ to $Q'$. This can be done 
by noting that 
\begin{eqnarray} 
Q' &=& \sigma_{j} Q \sigma_{k} \label{basischange1}
\end{eqnarray} 
where we have used $\sigma_{j} \sigma_{k} = -i \sigma_{i}$. Substituting Eq.\ \ref{basischange1} in Eq.\ \ref{qtran1} 
we find
\begin{eqnarray} 
Q_{\rm new} &=& (A Q - i B).(i C Q + D)^{-1} \label{transfinal} \\
A &=&  \tilde{a}_{\nu} \sigma_{j}, \quad B = \tilde{b}_{\nu} \sigma_{k} \nonumber\\
C &=& \tilde{c}_{\nu} \sigma_{j}, \quad D = \tilde{d}_{\nu} \sigma_{k} \nonumber
\end{eqnarray} 

The above transformations indicate that a square-pulse protocol involving different ${\rm SU}(1,1)$ subgroups 
do not have a $SU(1,1)$ valued $U(T,0)$; this can be seen by noting that one needs to carry out intermediate basis changes to find the 
final coordinate when the transformation involves different $U_{\mu}$  (as in Eqs.\ \ref{evol1} and \ref{evol2} of Sec.\ \ref{fidsec}) and a simple matrix product 
involving different $U_{\mu}$ does not yield the correct coordinate transformation. Our analysis here also demonstrates that the arrangement of coordinates in the 
initial quaternion matrix $Q$ may be done arbitrarily without affecting the final result as eluded to in Sec.\ \ref{qtn} }

\section{Conformal Algebra in $d$-dimensions}
\label{EConf_d}

In this appendix, for completeness, we briefly collect the well-known conformal algebra and the corresponding generators. We will closely follow the notations of \cite{francesco2012conformal}. The conformal algebra in $R^d$ is given by the SO$(d+1,1)$ algebra, with the following commutation relations:
\begin{eqnarray}
&& [D, P_\mu] = i P_\mu \ , [D, K_\mu] = -i K_\mu \ , [K_\mu, P_\nu] = 2i\left(  \delta_{\mu\nu} D-  M_{\mu\nu} \right)\ , \\
&& [M_{\mu\nu}, P_\alpha] = i(\delta_{\nu\alpha} P_\mu - \delta_{\mu\alpha} P_\nu ) \ , [M_{\mu\nu}, K_\alpha] = i(\delta_{\nu\alpha} K_\mu - \delta_{\mu\alpha} K_\nu ) \ , \\
&& [M_{\alpha\beta}, M_{\mu\nu}] = i (\delta_{\alpha\mu} M_{\beta\nu} + \delta_{\beta\nu} M_{\alpha \mu} - \delta_{\beta\mu} M_{\alpha \nu} - \delta_{\alpha \nu} M_{\beta \mu}) \ .
\end{eqnarray}
The differential operator representation of this algebra is given by
\begin{eqnarray}
&& P_\mu = - i \partial_\mu \ , D= -i x^\mu \partial_\mu \ , M_{\mu\nu} = - i(x_\mu \partial_\nu - x_\nu \partial_\mu) \ , \\
&& K_\mu = -2 i x_\mu x^\nu \partial_\nu + i x^2 \partial_\mu \ .
\end{eqnarray}
The corresponding Euclidean AdS spacetime can be described by the hyperboloid in $R^{d+1,1}$:
\begin{eqnarray}
- (X^0)^2 + (X^1)^2 + \ldots (X^{d+1})^2  = - L^2 \ .
\end{eqnarray}
In the global patch, the AdS metric is given by
\begin{eqnarray}
ds^2 = L^2 \left( \cosh^2 \rho\,\, d\tau^2 + d\rho^2 + \sinh^2 \rho\,\, dY_{d-1}^2\right) \ , \quad Y \cdot Y =1\ .
\end{eqnarray}
The SO$(d+1,1)$ algebra is generated by the following Killing vectors  in global coordinate:

\begin{align}
D &= -i \partial_\tau , \,\,\,  M_{\mu\nu} = -i \bigg( Y_\mu \tfrac{\partial}{\partial Y^\nu} -Y_\nu \tfrac{\partial}{\partial Y^\mu}  \bigg) , \nonumber \\
P_\mu &= -i e^{-\tau} \bigg[ Y_\mu ( \partial_\rho + \tanh \rho\,\, \partial_\tau ) + \frac{1}{\tanh \rho} \nabla_\mu \bigg], \,\,\, \nabla_\mu =\tfrac{\partial}{\partial Y^\mu}- Y_\mu Y^\nu\tfrac{\partial}{\partial Y^\nu}, \nonumber \\
K_\mu &= -i e^{\tau} \bigg[ Y_\mu ( -\partial_\rho + \tanh \rho\,\, \partial_\tau ) - \frac{1}{\tanh \rho} \nabla_\mu \bigg].
\end{align}
One can check that on the boundary $\rho=\infty$ these Killing vectors generate via Eq.\eqref{1-2a} the L\"uscher-Mack term $K_\mu + P_\mu$. 

\section{SL$(2,C)$, Quasi-primaries, Lorentz symmetry}
\label{CS_CFT}

In this appendix we collect some relevant formulae and basic statements regarding the SL$(2,C)$ action of the Lorentz group in $(3+1)$-dimensions. We will closely follow \cite{Pasterski:2016qvg, Pasterski:2017kqt, Crawley:2021ivb, Banerjee:2018gce}. It is well-known that given a four-vector $X^\mu$, we can associate a hermitian matrix:
\begin{eqnarray}
X = \begin{bmatrix}
	X^0 - X^3 & X^1 + i X^2  \\
	X^1 - i X^2 & X^0 + X^3 \\
	\end{bmatrix} \ , \quad {\rm det} (X) = - X^\mu X_\mu \ .
\end{eqnarray}
Now, an SL$(2,C)$ matrix $\Theta$ acts on the hermitian matrix $X$ as:
\begin{eqnarray}
X' = \Theta X \Theta^\dagger \ , \quad\Theta = \begin{bmatrix}
	a & b  \\
	c & d \\
	\end{bmatrix} \ , \quad {\rm with} \quad ad - bc =1\ ,
\end{eqnarray}
such that ${\rm det}(X') = {\rm det}(X)$.\footnote{In $(2+1)$-dimensions a similar description holds for a real-symmetric matrix\cite{Banerjee:2018gce}:
\begin{eqnarray}
X = \begin{bmatrix}
	X^0 - X^2 & X^1  \\
	X^1  & X^0 + X^2 \\
	\end{bmatrix} \ , \quad X' = \Lambda X \Lambda^T \ , \quad \Lambda \in SL(2,R) \ .
\end{eqnarray}
}

A more detailed action can be obtained by noting that the Lorentz group in $(3+1)$-dimensions acts as the global conformal group on the celestial sphere at infinity. More explicitly, given the Minkowski coordinates $X^\mu$, $\mu = 0, \ldots 3$, the celestial sphere is defined by $\eta_{\mu\nu} X^\mu X^\mu = 0$, on which we can define: $w = (X^1 + i X^2)/(X^0 + i X^3)$. Under a Lorentz transformation: $X'^\mu = \Lambda_\nu^\mu X^\nu$, we get: $w' = (a w+ b)/ (c w + d)$, where $ad - bc =1$ and $a,b,c,d$ are complex-valued. Subsequently, the $(3+1)$-dimensional scattering amplitudes can be expressed as celestial CFT correlators.

For an explicit realization, \cite{Pasterski:2016qvg, Pasterski:2017kqt, Crawley:2021ivb} has defined quasi-primary fields under this SL$(2,C)$ in the principal series, with a conformal dimension $\Delta = 1 + i s$, where $s$ is real-valued.\footnote{In $(d+1)$-dimensions, the corresponding conformal dimensions are $\Delta = \frac{d-1}{2} + i s$. } Moreover, there is a precise state-operator correspondence as well. The states are described by $\left|  h, \bar{h}, w, \bar{w}\right\rangle$ where $\{w, \bar{w}\}$ are the stereographic coordinates of the celestial sphere. Lorentz transformation, denoted by $U(\Lambda)$ acts on these states as follows:
\begin{eqnarray}
&& U(\Lambda) \left | h, \bar{h}, w, \bar{w} \right \rangle =  \frac{1} {(c w + d)^h} \frac{1} {(\bar{c} \bar{w} + \bar{d})^{\bar{h}}} \left | h, \bar{h} , \frac{a w + b} {c w+ d}, \frac{\bar{a} \bar{w} + \bar{b}}{\bar{c} \bar{w}+ \bar{d}} \right \rangle \ , \label{Ustate} \\
&& \Lambda \in SL(2,C) \ , \quad h = \frac{1 + i s - \sigma}{2} \ , \quad \bar{h} = \frac{1+ i s + \sigma}{2} \ ,
\end{eqnarray}
where $\sigma$ denotes the helicity of the massless particle.

The corresponding basis quasi-primary states are given by
\begin{eqnarray}
\left | s, \sigma, w=0=\bar{w} \right \rangle = \frac{1}{\sqrt{8\pi^4}} \int_0^\infty dE E^{i s} \left | E, 0, 0, E, \sigma \right \rangle \ ,
\end{eqnarray}
such that the states are normalized as follows:
\begin{eqnarray}
\left \langle s_1, \sigma_1, w_1, \bar{w}_1 | s_2, \sigma_2, w_2 , \bar{w}_2 \right \rangle  = \delta \left( s_1 - s_2\right)  \delta^{(2)} \left( w_1 - w_2\right)  \delta_{\sigma_1 \sigma_2} \ ,
\end{eqnarray}
where
\begin{eqnarray}
\left | s, \sigma, w , \bar{w} \right \rangle = \left( \frac{1}{1+ w\bar{w}} \right)^{1+ i s} U \left( R(w, \bar{w}) \right) \left | s, \sigma, 0 , 0 \right \rangle \ ,
\end{eqnarray}
in which $U \left( R(w, \bar{w}) \right)$ are the unitary rotation operators. The action of the Lorentz group on these states then yields equation (\ref{Ustate}), which essentially is comprised of SL$(2,C)$ matrix multiplications. The corresponding operators, denoted by ${\cal O}_{h, \bar{h}}$ will evolve in the Heisenberg picture according to:
\begin{eqnarray}
U(\Lambda) {\cal O}_{h, \bar{h}} \left( w, \bar{w} \right) U(\Lambda)^\dagger = \left( \frac{\partial \Lambda w}{\partial w}\right)^h \left( \frac{\partial \Lambda \bar{w}}{\partial \bar{w}}\right)^{\bar{h}} {\cal O}_{h, \bar{h}} \left( \Lambda w, \Lambda \bar{w}\right) \ .
\end{eqnarray}
A very similar description exists in $(2+1)$-dimensions as well, we refer the interested Reader to {\it e.g.}~\cite{Banerjee:2018gce} for more details on this.

\section{Fixed points and trajectories for a single $SU(1,1)$ subgroup}
\label{multiple}

For transformations which involve the subalgebra formed by $D, K_0, P_0$, one can choose
\ben
Q = x_0 I - i \sum_{i=1}^3 \sigma_i x_i
\label{5-2}
\een
so that the transformations on $R^4$ become (\ref{quatermobius}) with the entries $a_i$ real numbers satisfying (\ref{unitdeterminant}). In the following we will use the notation
\ben
\tilde{a} = \tilde{a}_1~~~~~\tilde{b} = i\tilde{a}_2~~~~~~~\tilde{c}=i\tilde{a}_3~~~~~~\tilde{d}=\tilde{a}_4
\een
so that the transformation is
\ben
Q^\prime = (\tilde{a} Q + \tilde{b}) (\tilde{c}Q + \tilde{d})^{-1}
\label{5-1}
\een
A fixed point $\bQ$ satisfies
\ben
\bQ (\tilde{c} \bQ + \tilde{d}) = \tilde{a}\bQ + \tilde{b}~~~~~~~~~\tilde{a}\tilde{d}-\tilde{b}\tilde{c}=1
\label{5-4}
\een
Using (\ref{5-4}) and standard properties of the Pauli matrices it is then easy to see that there are three kinds of fixed points or surfaces.
\begin{enumerate}

\item{} $ \bQ^\pm = \bx_0^\pm I $, i.e. $\bx_i = 0$. Here
\ben
\bx_0^\pm = \frac{1}{2\tilde{c}} \left[ (\tilde{a}-\tilde{d}) \pm \sqrt{(\tilde{a}+\tilde{d})^2 - 4} \right]
\label{5-5}
\een
In deriving (\ref{5-5}) we have used (\ref{5-1}). These are two fixed points and requires
\ben
(\tilde{a}+\tilde{d})^2 > 4
\label{5-6}
\een
The corresponding matrix
\ben
\begin{bmatrix} 
	\tilde{a} & \tilde{b} \\
	\tilde{c} & \tilde{d} \\
	\end{bmatrix}
	\quad
\label{5-7}
\een
is in the hyperbolic conjugacy class. 

\item{} When $(\tilde{a}+\tilde{d})^2 < 4$ there is a fixed two dimensional hypersurface defined by
\ben
\bx_0 = \frac{\tilde{a}-\tilde{d}}{2\tilde{c}}~~~~~\sum_{i=1}^3 (\bx_i)^2 = \frac{1}{4\tilde{c}^2} \left( 4 - (\tilde{a}+\tilde{d})^2 \right)
\label{5-8}
\een
and now the matrix in (\ref{5-7}) is in the elliptic conjugacy class.

\item{} The case $(\tilde{a}+\tilde{d})^2 = 4$ is marginal and corresponds to the parabolic conjugacy class.

\end{enumerate}

It is now straightforward to see that the transformation (\ref{5-1}) can be re-expressed in the following form for hyperbolic and elliptic transformations:
\ben
(Q^\prime - \bQ^+)(Q^\prime - \bQ^-)^{-1} = 
(\tilde{c} \bQ^- + \tilde{d}) (Q-\bQ^+) (Q-\bQ^-)^{-1}(\tilde{c} \bQ^+ + \tilde{d})^{-1}
\label{5-9}
\een
For the elliptic conjugacy class, the $\bQ^\pm$ in this equation refer to two antipodal points on the fixed 2-surface defined in (\ref{5-8}).

To prove (\ref{5-9}) we substitute (\ref{5-1}) in the left hand side of this equation so that the left hand side becomes
{\color{black}
\ben
\left[ (\tilde{a}-\tilde{c}\bQ^+)Q + (\tilde{b}-\tilde{d} \bQ^+) \right] \left[ (\tilde{a}-\tilde{c}\bQ^-)Q + (\tilde{b}-\tilde{d}\bQ^-) \right]^{-1}
\label{5-10}
\een
while the right hand side of (\ref{5-9}) becomes
\ben
\left[ (\tilde{c}\bQ^- +\tilde{d})Q - (\tilde{c}\bQ^-+\tilde{d})\bQ^+ \right] \left[  (\tilde{c}\bQ^+ +\tilde{d})Q - (\tilde{c}\bQ^+ +\tilde{d})\bQ^- \right]^{-1} 
\label{5-11}
\een
}
For the hyperbolic case, $\bQ^\pm$ are both proportional to the identity matrix. For the elliptic case, we we can take the antipodal points on the fixed 2-surface to be along the $3$ axis without any loss of generality. For both these cases one can then explicitly check that
\ben
\bQ^+ + \bQ^- = \frac{\tilde{a}-\tilde{d}}{\tilde{c}} \cdot I~~~~~\bQ^+\bQ^- = \bQ^- \bQ^+ = -\frac{\tilde{b}}{\tilde{c}} \cdot I
\label{5-12}
\een
Using these relations and the relation (\ref{5-4}) it may be checked that (\ref{5-10}) and (\ref{5-11}) are equal term by term.

{\color{black} The relationship (\ref{5-1}) can be iterated to yield
\begin{align}
Q'^{(n)} &= \bigg( {I} - ( \tilde{c}\, Q_- + \tilde{d})^n \alpha_\pm(Q) ( \tilde{c}\, Q_+ + \tilde{d})^{-n} \bigg)^{-1} \bigg( Q_+ - \left(( \tilde{c}\, Q_- + \tilde{d})^n \alpha_\pm(Q)  ( \tilde{c}\, Q_+ + \tilde{d})^{-n}\right) Q_- \bigg), \label{eq:ncycle}
\end{align}
where $$\alpha_\pm(Q) = \left[ (Q-Q_+)(Q-Q_-)^{-1} \right].$$
which can be used to determine the trajectory of a point under successive transformations.}

The Euclidean distance between two points represented by quaternions $Q_1$ and $Q_2$ are given by $\sqrt{|{\rm Det} (Q_1-Q_2) |}$. Equation (\ref{5-9}) implies
\ben
\frac{{\rm Det}(Q^\prime - \bQ^+)}{{\rm Det}(Q^\prime - \bQ^-)} = R^2 \frac{{\rm Det}(Q - \bQ^+)}{{\rm Det}(Q - \bQ^-)}
\label{5-13}
\een
where
\bea
R^2 & = & \frac{{\rm Det}(\tilde{c}Q^- + \tilde{d})}{{\rm Det}(\tilde{c}Q^+ + \tilde{d})}= \left(\frac{\tilde{a}+\tilde{d} -\sqrt{(\tilde{a}+\tilde{d})^2 -4}}{\tilde{a}+\tilde{d} +\sqrt{(\tilde{a}+\tilde{d})^2 -4}} \right)^2 \qquad ({\rm {Hyperbolic}}) \nonumber \\
R^2 & = &  1  \qquad \qquad  \qquad \qquad \qquad ({\rm{Elliptic}})
\label{5-14}
\eea
The equation can be recursively applied to yield
\ben
\frac{{\rm Det}(Q^{(n)} - \bQ^+)}{{\rm Det}(Q^{(n)} - \bQ^+)} = R^{2n} \frac{{\rm Det}(Q - \bQ^+)}{{\rm Det}(Q - \bQ^+)}
\label{5-13}
\een
where $Q^{n}$ denotes the transformed point after $n$ such transformations.
The expressions for $R^2$ then immediately implies that for hyperbolic transformations a point converges to one of the fixed points for large $n$, whereas for ellipic transformations, the point keeps moving on a circle of fixed radius around a point on the fixed surface. {Note that when we analytically continue to Lorentzian times we end up with $a = d^*$ which keeps the combination $a+d$ real. Hence we can classify dynamics based on if $R^2$ is a pure phase or real. This corresponds to $\text{Re}(a)$ being either less than 1 or greater. }

{\color{black} Using the relation (\ref{eq:ncycle}) we can calculate various physical quantities under a Floquet drive of this type, after $n$ cycles. }The result depends on the conjugacy class of the transformation at the end of a cycle. 
However the general features will be similar to the $d=1$ case, i.e. hyperbolic classes lead to a heating phase while elliptic classes lead to oscillatory phase \cite{Wen:2018agb, Wen:2020wee, Das:2021gts}.

When a cycle involves two different $SL(2,R)$ subgroups of the conformal group, e.g. two different $\Pi_\mu$'s, one can no longer represent the net transformation by a $SL(2,H)$ transformation with parameters proportional to the identity, {rather a change is basis in required, see Appendix \S\ref{appb}.The analysis of fixed points (surfaces) etc. needs to be re-done by investigating the conjugacy invariants of $SL(2,H)$ \cite{FOREMAN200425}, and the result of such Floquet drives will be quite different. } Explicit results for physical quantities under Floquet dynamics will appear in a future publication.





\end{document}